\documentstyle [12pt,a4,epsf] {article}

\newcommand{\e} { {\rm e} }
\renewcommand{\pd} {\partial}
\newcommand{\ie} {{\it i.e.}}
\newcommand{\eg} {{\it e.g.}}

\begin{document}

%-----------------------------------------------------------
% 1st page
%-----------------------------------------------------------

\title {Kinetics of Surfactant Adsorption \\
            at Fluid-Fluid Interfaces}
\author {Haim Diamant and David Andelman$^*$ \\
         \\
         School of Physics and Astronomy   \\
         Raymond and Beverly Sackler Faculty of Exact Sciences \\
         Tel Aviv University, Ramat Aviv, Tel Aviv 69978, Israel \\
         \\}
\date{April 1996}
\maketitle

%-----------------------------------------------------------
% Abstract
%-----------------------------------------------------------
\begin{abstract}
\setlength {\baselineskip} {20pt}

We present a theory for the kinetics of surfactant adsorption
at the interface between an aqueous solution and another fluid
(air, oil) phase.
The model relies on a free-energy formulation.
It describes both the diffusive transport of
surfactant molecules from the bulk solution to the interface, and
the kinetics taking place at the interface itself.
When applied to non-ionic surfactant systems, the theory recovers results
of previous models, justify their assumptions and predicts a 
diffusion-limited adsorption, in accord with experiments.
For salt-free ionic surfactant solutions, electrostatic interactions are shown
to drastically affect the kinetics.
The adsorption in this case is predicted to be kinetically limited, and
the theory accounts for unusual experimental results obtained recently 
for the dynamic surface tension of such systems.
Addition of salt to an ionic surfactant solution leads to screening of
the electrostatic interactions and to a diffusion-limited adsorption.
In addition, the free-energy formulation offers a general method for relating 
the dynamic surface tension to surface coverage. 
Unlike previous models, it does not rely on equilibrium relations
which are shown in some cases to be invalid out of equilibrium.

\end{abstract}

%-----------------------------------------------------------
% Introduction
%-----------------------------------------------------------

\pagebreak
\section{Introduction}
\label{introduction}
\setcounter{equation}{0}
\setlength {\baselineskip} {20pt}

Aqueous solutions of surface-active agents ({\em surfactants}) 
play a major role in various fields and applications, such as 
biological membranes, petrochemical processes, detergents, etc.
\cite{surfactant}.
In some important cases, equilibrium properties of the 
surfactant adsorption at interfaces are not sufficient, 
and knowledge of the kinetics is required.
Processes of fast wetting, foaming and stability of thin soap films
may serve as good examples.
The kinetics of surfactant adsorption have been addressed by
experimental and theoretical studies since the 1940s, and various 
experimental techniques have been devised, primarily aimed at the            
measurement of dynamic interfacial tensions \cite{review}.

The pioneering theoretical work of Ward and Tordai \cite{WT} 
considered a diffusive transport of surfactant molecules from a bulk
surfactant solution to an interface and formulated the time-dependent 
relation,
\begin{equation}
  \sigma(t) = \sqrt{\frac{D}{\pi}} \left[ 2c_b\sqrt{t} - \int_0^t
              \frac{c_1(\tau)}{\sqrt{t-\tau}} d\tau \right],
 \label{WToriginal}
\end{equation}
where $c_b$ is the bulk concentration and $D$ the surfactant diffusivity.
This equation gives only one relation between $\sigma(t)$, the surface 
density of surfactants adsorbed at the interface, and $c_1(t)$, the 
surfactant concentration at the sub-surface layer of solution.  

Subsequent theoretical research has focused on providing the second
closure relation between these two variables by introducing a certain
adsorption mechanism at the interface.
Various relations have been suggested, resembling equilibrium 
isotherms \cite{Delahay}--\cite{Joos1}, or having a
kinetic differential form \cite{Miller}--\cite{Franses}.
Such theories have been quite successful in describing the 
experimentally observed adsorption of common non-ionic surfactants.
Yet, they suffer from several drawbacks:
(i) the closure relation between the surface density and sub-surface
concentration, which expresses the kinetics taking place at the 
interface, is introduced as an {\em external} boundary condition, 
and does not uniquely arise from the model itself;
(ii) the calculated dynamic surface tension relies on 
an {\em equilibrium} equation of state, and assumes that it also 
holds out of equilibrium \cite{Fordham};
(iii) similar theories cannot be easily extended to describe more
complicated systems, such as {\em ionic} surfactant solutions
\cite{Langevin}.

In the current work we would like to present an alternative
approach to the kinetics of surfactant adsorption, overcoming 
these drawbacks.
In Sec.~\ref{non-ionic} we lay the foundations of our model \cite{letter}, 
based on a free-energy formulation. 
Non-ionic surfactants are considered, recovering results of previous
models and justifying their assumptions.
In particular, we show that the adsorption of non-ionic surfactants is
in general limited by diffusion from the bulk solution.
In Sec.~\ref{ionicnosalt} we modify the theoretical framework and apply 
it to salt-free ionic surfactant solutions.
A few models have been proposed for describing the kinetics
of ionic surfactant adsorption \cite{Dukhin}--\cite{Radke},
yet none of them is able to account for recent experimental results
for the dynamic surface tension of salt-free ionic surfactant
solutions \cite{Langevin}. 
We show that the adsorption in such systems is limited by the kinetic
processes at the interface. 
Consequently, we point out a problem common to all previous 
models. 
Using our model, we then account for the recent experimental 
findings.
Section~\ref{ionicwithsalt} considers ionic surfactant solutions with
added salt. The adsorption is shown in this case to be limited  
again by diffusion, and the effect of salt concentration is 
examined.
Finally, we present a few concluding remarks in Sec.~\ref{conclusion}
and point out possible future prospects.

%-----------------------------------------------------------
% Non-ionic Surfactants
%-----------------------------------------------------------
\section{Non-Ionic Surfactants}
\label{non-ionic}
\setcounter{equation}{0}

Consider an interface between an aqueous solution of non-ionic 
surfactants and an air or oil phase.
The system is schematically illustrated in Fig.~1.
We assume that the width of the interface is much smaller, and
its radius of curvature much larger, than any length scale relevant
to the adsorption process.
Hence, the interface can be regarded as sharp and flat, lying
at the plane $x=0$, and the problem is reduced to one dimension.
At $x \rightarrow \infty$, the solution is in contact with a 
bulk reservoir of surfactant molecules, where the chemical
potential and surfactant volume fraction are fixed to be 
$\mu_b$ and $\phi_b$, respectively.
We consider a dilute solution, \ie, the surfactant volume
fraction is much smaller than unity throughout the solution.
The concentration is also smaller than the critical
micelle concentration ({\it cmc}), so the surfactants are dissolved 
only as monomers.
At the interface itself, however, the volume fraction may become
large. 

%----------------------------------------------------
\vspace{0.5cm}
\subsection{The Free Energy}

We write the excess in free energy per unit area due to the 
interface (\ie, the change in interfacial tension), $\Delta\gamma$,
as a functional of the surfactant volume fraction in the bulk solution, 
$\phi(x>0)$, and its value at the interface, $\phi_0$,
\begin{equation}
  \Delta \gamma [\phi] = \int_0^\infty \Delta f[\phi(x)] dx + 
        f_0 (\phi_0).
 \label{Dg}
\end{equation}
The first term is the contribution from the bulk solution, $\Delta f$ 
being the excess in free energy per unit volume over the bulk, 
uniform state.
The second is the contribution from the interface itself, where
$f_0$ is the free energy per unit area of the surfactant at the 
interface.
The sharp, ``step-like" profile considered, has led us to treat the
bulk solution and the interface as two coupled sub-systems, 
rather than a single one \cite{Tsonop}.

The bulk sub-system is considered as an ideal, dilute solution, 
including only the ideal entropy of mixing and the contact with the 
reservoir and neglecting gradient terms, 
\begin{equation}
  \Delta f(\phi) = \frac{1}{a^3} \{ T [ \phi\ln\phi - \phi - 
    (\phi_b\ln\phi_b - \phi_b) ] - \mu_b 
    (\phi - \phi_b) \},
 \label{Df}
\end{equation}
where $a$ denotes the surfactant molecular dimension and $T$ the 
temperature (we set the Boltzmann constant to unity).

At the interface, however, since $\phi_0$ may become much larger 
than $\phi(x>0)$, we must take into account the finite molecular size 
and the interactions between surfactant molecules,
\begin{equation}
  f_0(\phi_0) = \frac{1}{a^2} \left\{ T [ \phi_0\ln\phi_0 + (1-\phi_0)
    \ln(1-\phi_0) ] - \alpha\phi_0 - \frac{\beta}{2}\phi_0^2 - 
    \mu_1\phi_0 \right\}.
 \label{f0}
\end{equation}
The term in the square brackets is the entropy of mixing, this time 
in its complete form, since $\phi_0$ is not necessarily small. 
The second accounts for the energetic preference of the surfactants to 
lie at the interface, $\alpha$ being positive by the definition of our 
molecules as surface-active.
The third is the energy of lateral interaction between neighboring 
surfactants at the interface, where $\beta$ is assumed to be positive 
too, \ie, expressing an overall {\em attractive} interaction.
The last term accounts for the contact with the solution adjacent
to the interface, where the chemical potential is 
$\mu_1\equiv\mu(x\rightarrow 0)$ \cite{barrier}.
 
Variation of $\Delta\gamma$ with respect to $\phi(x)$ yields the excess in
chemical potential at a distance $x$ from the interface,
\begin{eqnarray}
  \Delta \mu(x) &=& \mu(x)-\mu_b = 
                 a^2 \frac {\delta\Delta\gamma} {\delta\phi(x)} =
                 T \ln\phi(x) - \mu_b \ ; \ \ \ x > 0  
    \label{mux} \\
  \Delta \mu_0 &=& \mu_0-\mu_1 =
                 a^2 \frac {\delta\Delta\gamma} {\delta\phi_0} =
                 T \ln\frac{\phi_0}{1-\phi_0} - \alpha 
                 - \beta\phi_0 - \mu_1.
    \label{mu_0}
\end{eqnarray}
From Eq.~(\ref{mux}) we can deduce, as expected,
\begin{eqnarray}
  \mu_b &=& T \ln\phi_b  \nonumber \\
  \mu_1  &=& T \ln\phi_1,
\end{eqnarray}
where $\phi_1\equiv\phi(x\rightarrow 0)$ denotes the surfactant volume 
fraction at the sub-surface layer.
 
%----------------------------------------------------
\vspace{0.5cm}
\subsection{Thermodynamic Equilibrium}

In equilibrium, the chemical potential is equal to $\mu_b$ 
throughout the entire system (the variations of $\Delta\gamma$ vanish).
From Eq.~(\ref{mux}) we obtain the equilibrium profile,
\begin{equation}
  \phi(x) \equiv \phi_b \ ; \ \ \ x>0,
 \label{uniform}
\end{equation}
and from Eq.~(\ref{mu_0}), the equilibrium adsorption isotherm,
\begin{equation}
  \phi_0 = \frac {\phi_b} {\phi_b + 
           \e^{-(\alpha+\beta\phi_0)/T}}.
 \label{Frumkin}
\end{equation}
We have recovered the {\em Frumkin adsorption isotherm}, 
which reduces to the well known {\em Langmuir adsorption isotherm} 
\cite{Adamson} when the interaction term is neglected ($\beta=0$).
From Eqs.~(\ref{f0}), (\ref{mu_0}) and (\ref{uniform}) one 
also obtains the equilibrium equation of state,
\begin{equation}
  \Delta\gamma = \frac{1}{a^2} \left[ T\ln(1-\phi_0) + 
               \frac{\beta}{2} \phi_0^2 \right],
 \label{eqstate}
\end{equation}
which was previously derived from other, though equivalent 
considerations (integration of the Gibbs equation) \cite{Lin1}.

%----------------------------------------------------
\vspace{0.5cm}
\subsection{Out of Equilibrium}
\label{non-ionic_noneq}

Throughout our analysis we assume proportionality between 
velocities and the potential gradient \cite{Langer}, 
and take the surfactant mobility to be $D/T$ according to the 
Einstein relation ($D$ being the surfactant diffusivity).
At positions not adjacent to the interface this leads to the
following surfactant current density,
\begin{equation}
  j(x) = -\phi \frac{D}{T} \frac{\pd\mu}{\pd x} = 
         -D \frac{\pd\phi}{\pd x}.
 \label{jx}
\end{equation}
Applying the continuity condition, 
$\pd\phi/\pd t = -\pd j/\pd x$, 
we get the ordinary {\em diffusion equation},
\begin{equation}
  \frac{\pd\phi}{\pd t} = D \frac{\pd^2\phi}{\pd x^2}.
 \label{diffusion}
\end{equation}

The proximity of the interface requires a more careful treatment.
First, we discretize expression (\ref{Dg}) on a lattice with cells 
of size $a$,
\[
  \Delta \gamma[\phi] = a \sum_{i=1}^\infty \Delta f(\phi_i) + 
                        f_0(\phi_0),
\]
where $\phi_i \equiv \phi(x=ia)$. Discretized current densities, $j_i$,
can be similarly defined.
Since we do not allow molecules to leave the interface towards
the other (air, oil) phase (\ie, $j_0=0$), we have from the 
continuity condition
\[
  \frac{\pd\phi_0}{\pd t} = - \frac{j_1}{a},
\]
and can therefore write
\begin{equation}
  \frac{\pd\phi_1}{\pd t} = -\frac{j_2-j_1}{a} =
        \frac{D}{a} \left.
        \frac{\pd\phi}{\pd x} \right|_{x=a}
        - \frac{\pd\phi_0}{\pd t}.
 \label{dp1dt}
\end{equation}
Applying the Laplace transform to Eqs.~(\ref{diffusion}) and 
(\ref{dp1dt}) while assuming an initial uniform state, 
$\phi(x,t=0) \equiv \phi_b$, a relation is obtained between the 
surface and sub-surface volume fractions, $\phi_0$ and $\phi_1$,
\begin{equation}
  \phi_0(t) = \frac{1}{a} \sqrt{\frac{D}{\pi}} \left[
          2\phi_b\sqrt{t} - \int_0^t \frac {\phi_1(\tau)} 
          {\sqrt{t-\tau}} d\tau \right]
          + 2\phi_b - \phi_1.
 \label{WT}
\end{equation}
This relation is similar to the classical result of Ward and Tordai 
\cite{WT}, Eq.~(\ref{WToriginal}), except for the term 
$2\phi_b - \phi_1$. 
This difference is due to the fine details we have considered near
the interface and our initial condition. 
Ward and Tordai's analysis assumes a continuous profile up to
the non-aqueous phase, and hence replaces Eq.~(\ref{dp1dt})
with a simpler condition, 
$\pd\phi_0/\pd t = (D/a) \pd\phi/\pd x |_{x=0}$.
In addition, it requires an initial empty interface [$\phi_0(t=0)=0$],
whereas we set $\phi_0(t=0)=\phi_b$.
At any rate, the difference vanishes when the cell size, $a$, 
goes to zero.
Finally, we find the equation governing the kinetics at the interface 
itself \cite{differD},
\begin{equation}
  \frac{\pd\phi_0}{\pd t} = -\frac{j_1}{a} = 
          \frac{\phi_1}{a} \frac{D}{T}
          \frac{\mu_1-\mu_0}{a} = \frac{D}{a^2} \phi_1
          \left[ \ln \frac{\phi_1(1-\phi_0)}{\phi_0} + \frac{\alpha}{T}
          + \frac{\beta\phi_0}{T} \right].
 \label{dp0dt}
\end{equation}
Simultaneous solution of Eqs. (\ref{WT}) and (\ref{dp0dt}) yields
the solution of the adsorption problem, \ie, the time-dependent 
surface coverage, $\phi_0(t)$.

%------------------------------------------------------------
\vspace{0.5cm}
\subsection{Limiting Cases for the Adsorption}
\label{non-ionic_limit}

In writing the above equations, we have separated the kinetics of
the system into two coupled kinetic processes. 
The first takes place inside the bulk solution and is described by 
Eqs.~(\ref{diffusion}) and (\ref{dp1dt}) [or, alternatively, by
Eq.~(\ref{WT})],
whereas the second takes place at the interface and is described by 
Eq.~(\ref{dp0dt}).
Two important limiting cases correspond to the relative time scales 
of these two processes:
\begin{itemize}
 \item {\em Diffusion-limited adsorption (DLA)} applies when the 
  equilibration process inside the solution is much slower than 
  the one at the interface.
  One can then assume that the interface is in equilibrium at all times
  with the adjacent solution, \ie, the variation (\ref{mu_0})
  vanishes, and $\phi_0$ immediately responds to changes in $\phi_1$ via 
  the equilibrium isotherm.
 \item {\em Kinetically limited adsorption (KLA)} takes place when the 
  kinetic process at the interface is the slower one.
  In this case, the solution is assumed to be at all times in
  equilibrium with the bulk reservoir, \ie, the variation 
  (\ref{mux}) vanishes, 
  and $\phi_0$ changes with time according to Eq.~(\ref{dp0dt}).
\end{itemize}

One may suggest an alternative way of looking at the same limiting cases, 
the usefulness of which will become evident later on. 
Let us re-examine the expression for the interfacial contribution, 
$f_0$ [Eq.~(\ref{f0})]. 
The DLA case corresponds to the following
description (see Fig.~2a).
The interface is all the time at the minimum of the curve
$f_0(\phi_0)$, yet the shape of the curve changes with time as 
$\mu_1$ is changed by diffusion, until it attains the value of 
$\mu_b$. The surface coverage increases with time
as the minimum of $f_0$ is shifted to larger values of $\phi_0$.
On the other hand, KLA corresponds to a different scenario 
(Fig.~2b). The shape of $f_0(\phi_0)$ remains fixed since
$\mu_1$ is constantly equal to $\mu_b$. 
The surface coverage, $\phi_0$, increases until finally reaching the value
corresponding to the minimum of $f_0$.

In order to figure out whether one of the above limits applies to non-ionic
surfactant adsorption, the time scales of these two limiting cases must
be compared. 
Let us start with the DLA case and look for 
the asymptotic time dependence of the process.
We return to the Laplace transform of Eqs. (\ref{diffusion}) and 
(\ref{dp1dt}) and let the conjugate variable of the transform approach zero
\cite{Tauber}.
In addition, since the DLA limit is currently considered, we neglect the 
kinetics at the interface and take $\phi_0$ to be almost constant.
After inverting back to $t$-space, this procedure leads to the following 
asymptotic time dependence, previously mentioned by Hansen 
\cite{Hansen},
\begin{equation}
  \phi_b - \phi_1 \simeq \frac{a\phi_{0,eq}}{\sqrt{\pi D t}};
  \ \ \ \ \  t \rightarrow \infty,
 \label{asymp}
\end{equation} 
where $\phi_{0,eq}$ is the equilibrium surface coverage.
Looking at Eq.~(\ref{asymp}) we can identify the characteristic
{\em time scale of diffusion}, 
\begin{equation}
  \tau_d = \left( \frac{\phi_{0,eq}}{\phi_b} \right)^2 
        \frac{a^2}{D}.
 \label{td}
\end{equation}
Typical values of $a^2/D$ correspond to very short times (of
order $10^{-9}$ sec), but the prefactor of 
$(\phi_{0,eq}/\phi_b)^2$ is usually very large (of order, say, $10^{11}$). 
Thus, the diffusion time scales may reach minutes. 

In order to estimate the time scale of the kinetics taking place at the
interface, we look at the asymptotic behavior of Eq.~(\ref{dp0dt})
and find \cite{prevkinetic},
\begin{eqnarray}
  \phi_{0,eq} - \phi_0(t) &\sim& \e^{-t/\tau_k}
    \nonumber \\
  \tau_k &\simeq& \left( \frac{\phi_{0,eq}}{\phi_b} \right)^2
    \frac{a^2}{D} \e^{-(\alpha+\beta\phi_{0,eq})/T}.
%    {1-(\beta/T)\phi_{0,eq}(1-\phi_{0,eq})}
 \label{tk}
\end{eqnarray}
Since the value of $D$ at the interface is not expected to be 
drastically smaller than that inside the solution,
comparison of Eqs.~(\ref{td}) and (\ref{tk}) leads to the conclusion
that $\tau_d>\tau_k$. 
We thus expect, in general, that 
{\em non-ionic surfactants should exhibit diffusion-limited adsorption}.
This is one of the conclusions of our study as applied to non-ionic 
surfactants, and has been observed for quite a large number of such
surfactants \cite{review,latdiff}.
It is somewhat expected, since we did not include in the interfacial
free energy, Eq.~(\ref{f0}), any potential barrier that might lead 
to kinetic limitations.
The ``footprint" of DLA is the asymptotic
time dependence (see Eq.~(\ref{asymp}) and Ref.~\cite{Hansen})
\[ \phi_{0,eq}-\phi_0(t) \sim \sqrt{\tau_d/t}. \]
For any reasonable dependence between the surface tension and
surface coverage, $\gamma(t)=\gamma[\phi_0(t)]$, we expect the surface
tension to also exhibit a similar asymptotic time dependence,
$\gamma(t)-\gamma_{eq} \sim \sqrt{\tau_d/t}$.
It should be compared with the asymptotic {\em exponential} time dependence 
in the case of KLA [Eq.~(\ref{tk})].
Four examples of experimental results are given in Fig.~3,
all exhibiting an asymptotic $t^{-1/2}$ behavior, as expected for DLA. 

Having realized that the adsorption is diffusion-limited,
the mathematical solution of the problem amounts to the 
simultaneous solution of two equations:
(i) the Ward-Tordai equation, (\ref{WT}), accounting for
the diffusive transport;
(ii) an isotherm, such as (\ref{Frumkin}) (with $\phi_1$
replacing $\phi_b$), describing the (immediate) response of the surface 
coverage to changes in the sub-surface layer of solution.
Useful numerical schemes were suggested for performing such a task
\cite{review,Lin1}, yielding the time-dependent surface coverage, 
$\phi_0(t)$.

%-----------------------------------------------------------
\vspace{0.5cm}
\subsection{Dynamic Surface Tension in a DLA Process}

Since most experiments measure dynamic surface tensions
and not surface coverages, we still need a relation between
these two variables in order to relate theoretical calculations 
to actual measurements.
We return, therefore, to the evolution of the interfacial tension during a
DLA process.
As was stated above, in this limit the interfacial contribution, 
$f_0(\phi_0)$, is all the time at its minimum.
We can, therefore, write
\[
  \Delta \gamma [\phi] = \int_0^\infty \Delta f[\phi(x)] dx + 
        \frac{1}{a^2} \left[ T\ln(1-\phi_0) + \frac{\beta}{2} \phi_0^2 
        \right].
\]
If, in addition, the contribution from the bulk solution is neglected
(recalling that it completely vanishes when equilibrium is reached),
we are left with the equilibrium relation, Eq.~(\ref{eqstate}).
Hence, the equilibrium equation of state, relating the surface 
tension to the surface coverage, holds approximately also
out of equilibrium.
Note, that this conclusion is valid only in the case of 
{\em diffusion-limited adsorption}.

Practically all previous works assumed that the equilibrium relation
between the surface tension and surface coverage holds for 
the {\em dynamic} surface tension as well.
As we have just found, non-ionic surfactants usually do undergo 
DLA.  
Hence, the assumption employed by previous works was
justified, as far as non-ionic surfactants were concerned.
Indeed, satisfactory agreement with experimental findings is
obtained, when results of the numerical schemes mentioned above
are related to the dynamic surface tension via the equilibrium 
equation of state \cite{Lin1,Lin2}.
However, this conclusion is drastically modified for {\em ionic}
surfactants, as we show in the next section.

The dependence of $\Delta\gamma$ on $\phi_0$, as defined by 
Eq.~(\ref{eqstate}), is shown in Fig.~4a.
Note the moderate slope in the beginning of the process;
the surface coverage significantly changes without a corresponding
change in the surface tension. 
It is a result of the competition between the entropy and interaction 
terms in Eq.~(\ref{eqstate}).
As the surface coverage increases, the surface tension starts falling
until reaching its equilibrium value.
Since $\phi_0$ monotonically increases with time during the adsorption, 
we expect the {\em time} dependence of $\Delta\gamma$ to resemble the
schematic curve depicted in Fig.~4b: a slow change 
in the beginning, then a rapid drop, and eventually a relaxation towards
equilibrium.
This is, indeed, in agreement with dynamic surface tension measurements
(\eg, \cite{Lin1}).
Returning to Fig.~4a, the surface tension will start its rapid fall
roughly when the second derivative of $\Delta\gamma$ with respect to $\phi_0$ 
changes sign, \ie, when
\[ 1-\phi_0 \sim (\beta/T)^{-1/2}. \]
As one examines surfactant solutions of increasing bulk concentrations 
(but always below the {\it cmc}), the surface coverage corresponding to the 
beginning of the drop in surface tension will be reached earlier along the process.
The initial period of slow change in the tension will shrink, 
until finally vanishing behind the finite experimental resolution.
This trend is observed experimentally \cite{Lin2}.

The need for an interaction between surfactant molecules in order to
account for such a time dependence of surface tensions was previously
realized by Lin {\it et al.} \cite{Lin2}. 
They even suggested the existence of a transition from a gaseous to a
liquid phase as being responsible for the initial period of almost constant
tension.
In fact, the form of $f_0$ as defined in Eq.~(\ref{f0}) may account
for a two-phase region, but only if $\beta>4T$.
As demonstrated above in Fig.~4 (where $\beta=3T$), 
this is not necessary, however, for recovering the experimentally 
observed time dependence.

%---------------------------------------------------------------
% Ionic Surfactants without Salt
%---------------------------------------------------------------
\section{Ionic Surfactants without Added Salt}
\label{ionicnosalt}
\setcounter{equation}{0}

We now consider the problem of {\em ionic} surfactant adsorption.
The main difference compared to the previous, non-ionic case
is the introduction of electrostatic interactions.
The kinetics of the system include, apart from the diffusive transport of
molecules and their adsorption at the interface, the formation
of an electric double layer due to the increasing surface charge.
We start with the case of a salt-free solution, where the only charges
present are those of the surfactant ions and their balancing 
counter-ions. 
The system is schematically illustrated in Fig.~5.
In such a case the electrostatic interactions are unscreened, 
and thus, as we shall see, have a drastic effect on the adsorption 
process.
Instead of the single degree of freedom we have specified in 
the non-ionic case, namely the surfactant profile, $\phi(x,t)$, 
we should consider here three degrees of freedom: 
the surfactant ion profile, $\phi^+(x,t)$, the counter-ion profile, 
$\phi^-(x,t)$, and the local mean electric potential, $\psi(x,t)$
(without loss of generality, we take the surfactant ions as the 
positive ones).

For simplicity, we assume the following:
(i) the surfactant molecules are fully ionized; 
(ii) the surfactant ions and counter-ions are monovalent 
(extension to general valencies, however, is quite straightforward
within our mean-field formulation);
(iii) as in the previous section, the solution is assumed to be dilute.
In the current context this last assumption also allows us to treat the 
surfactant and counter-ion volume fractions as independent variables,
as the probability for an ion and a counter-ion to overlap at the
same spatial position is negligible.
In addition, and without loss of generality, we take the surfactant ions 
to be the positive ones.

The extension of the model to ionic surfactants evidently demonstrates
the benefits of the free-energy formulation we have employed.
We just repeat the scheme of Sec.~\ref{non-ionic} while adding terms for
the counter-ions and electrostatic interactions.
Unlike previous models and due to its simplicity, this formulation will 
allow us to clarify the complex problem of ionic surfactant adsorption
and reach novel conclusions concerning it.

%----------------------------------------------------------
\vspace{0.5cm}
\subsection{The Free Energy}

Following the analysis of Sec.~\ref{non-ionic}, we write the change in
interfacial tension in the form
\begin{equation}
  \Delta \gamma [\phi^+,\phi^-,\psi] = \int_0^\infty 
        \{ \Delta f^+[\phi^+] 
        + \Delta f^-[\phi^-] + f_{el}[\phi^+,\phi^-,\psi] \} dx + 
        f_0 (\phi^+_0,\psi_0).
 \label{Dg2}
\end{equation}
The first two terms are the contributions from the bulk solution, 
depending on the surfactant and counter-ion volume fractions.
The third term is the electrostatic energy stored in the bulk solution.
The last term is the contribution from the interface itself, depending
on the interfacial values of the surfactant volume fraction, $\phi^+_0$, and 
the electric potential, $\psi_0$.
The counter-ions are assumed to be surface-inactive, and their
contribution to this term is neglected \cite{counterads}. 
As in the previous discussion, the two types of ions dissolved in the 
bulk solution have contributions coming from their entropy of mixing
and chemical potentials,
\begin{equation}
  \Delta f^\pm(\phi^\pm) = \frac{1}{(a^\pm)^3} \{ T [ 
    \phi^\pm\ln\phi^\pm - \phi^\pm - 
    (\phi^\pm_b\ln\phi^\pm_b - \phi^\pm_b) ] - 
    \mu^\pm_b (\phi^\pm - \phi^\pm_b) \},
 \label{Df2}
\end{equation}
where $a^\pm$ are the molecular dimensions of the two ions
and $\phi^\pm_b$, $\mu^\pm_b$ 
their volume fractions and chemical potentials, respectively, at the 
bulk reservoir.
Our assumption of monovalent ions implies
\[ \frac{\phi^+_b}{(a^+)^3} = \frac{\phi^-_b}{(a^-)^3} = 
   c_b,\]
where $c_b$ is the bulk concentration.   

The electrostatic term contains the interaction between the ions and
the electric field and the energy associated with the field
itself,
\begin{equation}
  f_{el} = e \left( \frac{\phi^+}{(a^+)^3} - \frac{\phi^-}{(a^-)^3} 
           \right) \psi - \
           \frac{\epsilon}{8\pi} \left( \frac{\pd\psi}
           {\pd x} \right)^2,
 \label{fel}
\end{equation}
where $e$ is the electronic charge and $\epsilon$ the dielectric 
constant of the solvent (water).
Finally, the modified expression for the interfacial contribution, 
$f_0$, is obtained from Eq.~(\ref{f0}) just by adding an electrostatic term,
\begin{equation}
  f_0(\phi^+_0,\psi_0) = \frac{1}{(a^+)^2} \left\{ T [
    \phi^+_0\ln\phi^+_0 + (1-\phi^+_0)
    \ln(1-\phi^+_0) ] - \alpha\phi^+_0 - \frac{\beta}{2}(\phi_0^+)^2  
     - \mu^+_1\phi^+_0 + e\phi^+_0\psi_0 \right\} .
 \label{f02}
\end{equation}

When the interface is still almost uncharged, another electrostatic
contribution should be considered, namely that of the interaction 
between the ions and their ``image" charges beyond the interface, 
as was pointed out by Onsager and Samaras \cite{Onsager}. 
This force decays like $1/x^2$, and since the dielectric constant of 
the other phase (air, oil), $\epsilon_o$, is smaller than that of the 
solvent, $\epsilon$, it is repulsive.
The ``image'' force is comparable to the repulsion from the surface 
charge for distances 
\[
  x < \sqrt{ \frac{1}{16\pi\phi^+_0} \frac{\epsilon-\epsilon_o}
      {\epsilon+\epsilon_o} } a^+.
\]
Obviously, when the right-hand side of this expression becomes smaller 
than the molecular dimension, $a^+$, this force is irrelevant.
For a water-air or water-oil interface this happens when
$\phi^+_0 \sim 0.02$, \ie, very soon along the adsorption 
process.
We allow ourselves, therefore, to disregard such early stages of
adsorption and neglect this interaction altogether
(it should be mentioned, however, that in the case of surface-{\em inactive}
electrolytes, the ``image'' interaction has a significant contribution to the 
interfacial tension).

We now take the variation of $\Delta\gamma$ with respect to $\phi^\pm(x)$ 
to get the excess in electrochemical potentials,
\begin{eqnarray}
  \Delta \mu^\pm(x) &=& \mu^\pm(x)-\mu^\pm_b =
         (a^\pm)^2 \frac {\delta\Delta\gamma} {\delta\phi^\pm} = 
         T \ln\phi^\pm \pm e\psi - \mu^\pm_b \ ; \ \ \ x>0  
    \label{mux2} \\
  \Delta \mu^+_0 &=& \mu^+_0-\mu^+_1 =
                 (a^+)^2 \frac {\delta\Delta\gamma} {\delta\phi^+_0} =
                 T \ln\frac{\phi^+_0}{1-\phi^+_0} - \alpha 
                 - \beta\phi^+_0 + e\psi_0 - \mu^+_1.
    \label{mu_02}
\end{eqnarray}  
We require that the electric potential vanish far away from the interface,
that is, $\psi(x\rightarrow\infty) = 0$,
and hence, deduce from Eq.~(\ref{mux2}), as expected,
\begin{eqnarray}
  \mu^\pm_b &=& T \ln\phi^\pm_b  \nonumber \\
  \mu^\pm_1  &=& T \ln\phi^\pm_1 \pm e\psi_1,
\end{eqnarray}
where $\phi^+_1$ and $\psi_1$ denote, respectively, the sub-surface 
values of the surfactant volume fraction and electric potential.
When we take the variation of $\Delta\gamma$ with respect to $\psi(x)$
and set it to zero (since only electrostatic effects are considered), the
{\em Poisson equation} is obtained,
\begin{equation}
  \frac{\pd^2\psi}{\pd x^2} = -\frac{4\pi e}{\epsilon}
  \left( \frac{\phi^+}{(a^+)^3} - \frac{\phi^-}{(a^-)^3} \right).
 \label{Poisson}
\end{equation}
Finally, variation with respect to $\psi(x=0)\equiv\psi_0$ yields
the expected boundary condition,
\begin{equation}
  \left. \frac{\pd\psi}{\pd x} \right|_{x=0} =
       -\frac{4\pi e}{\epsilon (a^+)^2} \phi^+_0,
 \label{neutral}
\end{equation}
which is equivalent to the requirement of overall charge neutrality 
(only surfactant ions, and no counter-ions, are allowed to be 
adsorbed at the interface).

%--------------------------------------------------------
\vspace{0.5cm}
\subsection{Thermodynamic Equilibrium} 

In equilibrium, the variations (\ref{mux2}) and (\ref{mu_02})
vanish, and we recover the {\em Boltzmann distributions},
\begin{equation}
  \phi^\pm(x) = \phi^\pm_b \e^{\mp e\psi(x)/T}; \ \ \ \ \ x>0,
 \label{Boltzmann}
\end{equation}
and the adsorption isotherm,
\begin{equation}
  \phi^+_0 = \frac {\phi^+_b} {\phi^+_b + 
           \e^{-(\alpha+\beta\phi^+_0-e\psi_0)/T}}.
 \label{Davies}
\end{equation}
We have recovered the {\em Davies adsorption isotherm} for ionic 
surfactants \cite{Davies}.
Combining Eqs. (\ref{Poisson}) and (\ref{Boltzmann}) leads to the 
well known {\em Poisson-Boltzmann equation},
\begin{equation}
  \frac{\pd^2\psi}{\pd x^2} = \frac{8\pi e c_b}{\epsilon}
           \sinh \frac{e\psi}{T},
 \label{PB}
\end{equation}
which determines the equilibrium double-layer potential 
\cite{VO,AndelmanES}.
Integrating this equation once and using the boundary condition 
(\ref{neutral}) yield the surface potential,
\begin{equation}
  \psi_0 = \frac{2T}{e} \sinh^{-1}(\lambda\phi^+_0),
 \label{psi0}
\end{equation}
where
\[ \lambda \equiv \frac{\kappa a^+}{4\phi^+_b} \]
and $\kappa^{-1} \equiv (8\pi c_b e^2/\epsilon T)^{-1/2}$ is the
{\em Debye-H\"{u}ckel screening length}.
Substituting $\psi_0$ from Eq.~(\ref{psi0}), the Davies isotherm 
can be expressed in terms of $\phi^+_0$ alone,
\begin{equation}
  \phi^+_0 = \frac {\phi^+_b} {\phi^+_b + [ \lambda\phi^+_0 + \sqrt{
           (\lambda\phi^+_0)^2+1} ]^2
           \e^{-(\alpha+\beta\phi^+_0)/T}}.
 \label{Davies2}
\end{equation}
From Eqs.~(\ref{Boltzmann}), (\ref{PB}) and (\ref{neutral}) one can
calculate the contribution of the bulk solution to the interfacial
free energy at equilibrium,
\begin{eqnarray}
  f_{bulk}^{eq} &\equiv& \int_0^\infty (\Delta f^+ + \Delta f^- 
                    + f_{el}) dx \nonumber \\
           &=& -\frac{2T}{(a^+)^2 \lambda} \left( \cosh \frac{e\psi_0}{2T}
               - 1 \right) =
               -\frac{2T}{(a^+)^2 \lambda} [\sqrt{(\lambda\phi^+_0)^2+1} - 1].
 \label{fbulk}
\end{eqnarray}
From Eqs.~(\ref{f02}), (\ref{mu_02}) and (\ref{fbulk}) we then get
the {\em equilibrium equation of state},
\begin{equation}
  \Delta\gamma = \frac{1}{(a^+)^2} \left[ T\ln(1-\phi^+_0) + 
               \frac{\beta}{2} (\phi^+_0)^2 - \frac{2T}{\lambda} 
               (\sqrt{(\lambda\phi^+_0)^2+1} - 1) \right].
 \label{eqstate2}
\end{equation}
In the limit of weak electric fields ($\lambda\phi^+_0 \ll 1$), 
the electrostatic repulsion between surfactant ions at the interface 
predominates. 
As a result, the electrostatic correction to the non-ionic equation of state 
(\ref{eqstate}) is quadratic in $\phi^+_0$ and effectively reduces the lateral 
attraction term.
In the limit of strong electric fields ($\lambda\phi^+_0 \gg 1$),
however, the high concentration of counter-ions near the interface 
makes the electrostatic term become only linear in $\phi^+_0$.

%-----------------------------------------------------
\vspace{0.5cm}
\subsection{Out of Equilibrium}

We write the current densities as in the previous section,
\begin{equation}
  j^\pm(x) = -\phi^\pm \frac{D^\pm}{T} \frac{\pd\mu^\pm}{\pd x} = 
         -D^\pm \left( \frac{\pd\phi^\pm}{\pd x} \pm \frac{e}{T}
         \phi^\pm \frac{\pd\psi}{\pd x} \right),
 \label{jx2}
\end{equation}
where $D^\pm$ are the diffusivities of the two ions.
Applying the continuity condition, the {\em Smoluchowski 
diffusion equations} are obtained,
\begin{equation}
  \frac{\pd\phi^\pm}{\pd t} = D^\pm \frac{\pd}{\pd x}
      \left( \frac{\pd\phi^\pm}{\pd x} \pm \frac{e}{T}
      \phi^\pm \frac{\pd\psi}{\pd x} \right).
 \label{diffusion2}
\end{equation}
 
As in Sec.~\ref{non-ionic}, we treat separately the positions adjacent
to the interface by discretizing the expressions for the various 
contributions and considering the current densities
near the interface.
The condition (see Sec.~\ref{non-ionic_noneq})
\[
   \frac{\pd\phi^+_0}{\pd t} = -\frac{j^+_1}{a^+}
\]
leads in this case to
\begin{equation}
  \frac{\pd\phi^\pm_1}{\pd t} = \frac{D^\pm}{a^\pm} \left(
        \left.\frac{\pd\phi^\pm}{\pd x}\right|_{x=a^\pm} \pm 
        \frac{e}{T} \phi^\pm_1 \left.\frac{\pd\psi}{\pd x}
        \right|_{x=a^\pm} \right) - \frac{\pd\phi^\pm_0}{\pd t}.
 \label{dp1dt2}
\end{equation}
The kinetic equation for the surfactant adsorption at the interface 
turns out to be
\begin{equation}
  \frac{\pd\phi^+_0}{\pd t} = 
          \frac{D^+}{(a^+)^2} \phi^+_1
          \left[ \ln \frac{\phi^+_1(1-\phi^+_0)}{\phi^+_0} + 
          \frac{\alpha}{T} + \frac{\beta\phi^+_0}{T} 
          -\frac{e(\psi_0-\psi_1)}{T} \right],
 \label{dp0dt2}
\end{equation}
where the last term in the brackets may be viewed as an electrostatic 
barrier located at the edge before the interface.
It can be also written, by means of Eq.~(\ref{neutral}), as
\[
  \frac{e(\psi_0-\psi_1)}{T} \simeq \frac{4\pi l}{a^+} \phi^+_0,
\]
where $l \equiv e^2/\epsilon T$ is the Bjerrum length (about 7 \AA \ 
for water at room temperature).
The kinetic equation can then be expressed, like its non-ionic parallel,
Eq.~(\ref{dp0dt}), in terms of $\phi^+_0$ and $\phi^+_1$ alone.
Finally, since we have assumed negligible adsorption of counter-ions 
at the interface, we may require
\begin{equation}
  \frac{\pd\phi^-_0}{\pd t} = 0.
 \label{nocounter}
\end{equation}

The Smoluchowki equations (\ref{diffusion2}) together with the Poisson
equation (\ref{Poisson}) make a set of three differential equations for
the three unknown functions, $\phi^+(x,t)$, $\phi^-(x,t)$ and $\psi(x,t)$.
Equations (\ref{neutral}), (\ref{dp1dt2}), (\ref{dp0dt2}) and 
(\ref{nocounter}) set the boundary conditions for these functions at
the interface. 
If we add appropriate boundary conditions at infinity (where the 
volume fractions converge to their bulk value and the electric 
potential vanishes) and initial conditions (say, a perfectly uniform initial 
state with a vanishing electric potential), the mathematical problem is 
well posed and, at least in principle, solvable.
The task of solving this system of equations, nevertheless, seems rather
formidable, though similar systems have recently been dealt with
numerically by MacLeod and Radke \cite{Radke}.
Fortunately enough, in all practical cases one can avoid the 
elaborate mathematical treatment, as will be demonstrated below.

%-----------------------------------------------------
\vspace{0.5cm}
\subsection{Limiting Cases for the Adsorption}

As in Sec.~\ref{non-ionic}, we are interested again in the distinction
between {\em diffusion-limited adsorption}, where the kinetics are 
controlled by the diffusive transport inside the solution, and 
{\em kinetically limited adsorption}, where the process is controlled
by the kinetics at the interface.
The kinetics inside the solution are governed now by 
Eqs.~(\ref{diffusion2}). In order to identify the corresponding time 
scale, let us assume for simplicity that the ions have equal
diffusivities, $D^+=D^-\equiv D$, and that the electric field is
weak, $e\psi/T\ll 1$. 
If we add the two equations (\ref{diffusion2}) and recall the 
Poisson equation (\ref{Poisson}), the sum $\phi^+(x,t) + \phi^-(x,t)$ 
is found, to first order in the electric field, to undergoes a 
{\em free} diffusion.
Since both at $t=0$ and $t\rightarrow\infty$ this sum is
$\phi^+_b+\phi^-_b$ for any $x$, it follows that it remains  
unchanged during the entire process.
Keeping this conclusion in mind and again making use of the 
Poisson equation (\ref{Poisson}), we now subtract the two equations
(\ref{diffusion2}) and obtain a linear diffusion equation for the electric 
potential, $\psi$,
\begin{equation}
  \frac{\pd\psi}{\pd t} = D \left( \frac{\pd^2\psi}
  {\pd x^2} - \kappa^2 \psi \right).
 \label{DLdiffusion}
\end{equation}
This equation describes the kinetics of relaxation of the electric
double layer. Its characteristic length scale is, as expected, the
Debye-H\"{u}ckel screening length, $\kappa^{-1}$, and the resulting
time scale is
\begin{equation}
  \tau_e = (\kappa^2 D)^{-1}.
 \label{te}
\end{equation}
More rigorously, we can calculate the asymptotic solution of
Eq.~(\ref{DLdiffusion}), obtaining the following interesting
time dependence,
\begin{equation}
  \psi(x,t\rightarrow\infty) \simeq \psi_0 \left( t - \frac{x}{2\kappa D}
       \right) \Theta \left( t - \frac{x}{2\kappa D} \right) \e^{-\kappa x},
 \label{psi_asymp}
\end{equation}
where $\Theta(t)$ is the step (Heaviside) function.
This potential may be viewed as a ``retarded equilibrium potential'':
a point $x$ in the solution is in some sense in equilibrium with the
interface (hence the exponential profile), but with the interface as it
was some time ($x/2\kappa D$) ago.
The information propagates from the interface to the solution with 
the velocity $2\kappa D$.
From Eq.~(\ref{psi_asymp}) it is clearly verified that the length
scale is $\kappa^{-1}$, and the time scale is $\tau_e$ as
defined in Eq.~(\ref{te}).

Equation~(\ref{te}) states the time scale needed for the electric double 
layer to adjust and attain equilibrium with the interface. 
Yet, in a salt-free solution, the surfactant and counter-ion profiles 
themselves construct the electric double layer, and therefore, $\tau_e$
is also the time scale of relaxation of the surfactant profile.

Typical values for $D$ are about $10^{-6}$ cm$^2$/sec, and 
$\kappa^{-1}$ in salt-free ionic surfactant solutions amounts to
hundreds of Angstroms.
This yields very small values for $\tau_e$ (on the order of 
microseconds). 
The relaxation of the profile, therefore, is much faster 
in the case of ionic surfactants (without added salt) than 
in the case of non-ionic ones.
This effect is due to the strong electrostatic interactions which 
drastically accelerate the kinetics inside the solution. 
In other words, the diffusion inside a salt-free ionic surfactant 
solution is an {\em ambipolar diffusion} \cite{LL} rather than a
regular one.
If the ion diffusivities are not assumed to be equal, one should 
replace $D$ in Eq.~(\ref{te}) with some effective diffusivity.
If the electric field is not weak, as is practically always the case in 
salt-free surfactant solutions, then the time scale of profile relaxation 
will be even {\em shorter}, and the above conclusion, obtained for
weak electric fields, will not change.

Turning to the kinetics which take place at the interface, we treat
Eq.~(\ref{dp0dt2}) similar to Eq.~(\ref{dp0dt}) of Sec.~\ref{non-ionic}, 
expand it close to equilibrium and find the time scale,
\begin{eqnarray}
  \tau_k &\simeq& \left( \frac{\phi^+_{0,eq}}{\phi^+_b} \right)^2
    \frac{(a^+)^2}{D^+} \exp \{-[\alpha+\beta\phi_{0,eq}
    -e(\psi_{0,eq}+\psi_{1,eq})]/T\} \nonumber \\
             &=& \tau_k^{(0)} \exp [e(\psi_{0,eq}+\psi_{1,eq})/T],
 \label{tk2}
\end{eqnarray}
where $\tau_k^{(0)}$ denotes the kinetic time scale found in the absence
of electrostatics [Eq.~(\ref{tk})].
As expected, the electrostatic repulsion of surfactant ions from
the charged interface slows down the adsorption process. 
The sum $\psi_0+\psi_1$ may also be written as $2\psi_1+(\psi_0-\psi_1)$,
where the first term expresses the slowing down due to the 
sub-surface concentration (which is lowered because of $\psi_1$), 
and the second accounts for a further slowing effect due to the edge 
electrostatic barrier.
If the electric field is strong, as is practically the case in
salt-free surfactant solutions, the duration of the process may become 
longer by orders of magnitude. This can be verified experimentally when 
the electrostatic interactions are screened by added salt 
\cite{Langevin}.
Using Eq.~(\ref{psi0}) in the limit of strong fields, together with 
Eq.~(\ref{neutral}), we can estimate the slowing factor by
\begin{equation}
  \exp [e(\psi_{0,eq}+\psi_{1,eq})/T] \simeq 
    \left( \frac{\kappa a^+}{2} \frac{\phi^+_{0,eq}}{\phi^+_b} 
    \right)^4 \exp \left( -\frac{4\pi l}{a^+} \phi^+_{0,eq} \right).
\end{equation}
This factor is typically very large. For example, in the experimental
system of Ref.~\cite{Langevin} one finds $\kappa a^+ \sim 10^{-2}$, 
$\phi^+_{0,eq}/\phi^+_b \sim 10^5$ and $l/a^+ \sim 1$, so the 
slowing factor amounts to about $10^7$ \cite{reservation}.

We see that the strong electrostatic interactions present in salt-free
ionic surfactant solutions drastically shorten the time scale of diffusion
inside the solution and drastically lengthen the time scale of kinetics
at the interface. 
We expect, therefore, that 
{\em ionic surfactants in salt-free solutions should exhibit kinetically 
limited adsorption}.
On one hand this conclusion greatly simplifies the mathematical
treatment of the problem: we can safely assume that the electric double 
layer is in quasi-equilibrium with the changing interface 
(\ie, it obeys the Poisson-Boltzmann theory) and deal with the 
kinetics at the interface alone.
On the other hand, the conclusion of kinetic control invalidates some
of the assumptions employed by previous models. It implies that the 
relevant adsorption scenario of the two described in Sec.~\ref{non-ionic_limit}
is the second one, illustrated in Fig.~2b.
We recall that according to this limiting scenario, the interfacial contribution 
to the free energy, $f_0(\phi^+_0)$, retains the same shape throughout the 
process and reaches its minimum only at equilibrium.
Hence, one cannot use in this case the equilibrium equation of state,
such as Eq.~(\ref{eqstate2}), to calculate dynamic surface tensions.

Having realized that the adsorption is limited by the kinetics at the
interface, we can take the ion profiles and electric double layer to be in
quasi-equilibrium. 
The Poisson-Boltzmann theory, therefore, can be implemented, and the 
kinetic equation, (\ref{dp0dt2}), is rewritten as
\begin{equation}
  \frac{\pd\phi^+_0}{\pd t} = \left(\frac{D^+\phi^+_b}{(a^+)^2}\right) \frac
      {\exp [(4\pi l/a^+)\phi^+_0]} {[\lambda\phi^+_0+\sqrt{(\lambda\phi^+_0)^2+1}]^2}
       \left\{ \ln \left[ \frac{\phi^+_b(1-\phi^+_0)}{\phi^+_0} \right] +
      \frac{\alpha}{T} + \frac{\beta\phi^+_0}{T} - 2\sinh^{-1}(\lambda\phi^+_0) \right\}
 \label{dp0dt2_kin}
\end{equation}
The mathematical problem of finding the time-dependent surface coverage, 
$\phi^+_0(t)$, is thus reduced to a single integration.

%------------------------------------------------------------
\vspace{0.5cm}
\subsection{Dynamic Surface Tension in a KLA Process}

In order to relate calculated surface coverages to measured surface tensions,
we need, as in Sec.~\ref{non-ionic}, an appropriate relation between these 
two variables.
As we have just concluded, the equilibrium equation of state will not
do in this case.
Since kinetically limited adsorption is considered, the contribution to the free 
energy from the bulk solution has the equilibrium dependence on the surface 
coverage, \ie, $f_{bulk}^{eq}(\phi^+_0)$ of Eq.~(\ref{fbulk}).
Hence, we can write the free energy (or equivalently, the dynamic
surface tension) as a function of the surface coverage alone,
\begin{eqnarray}
  \Delta\gamma[\phi^+_0(t)] &=& f_0(\phi^+_0) + f_{bulk}^{eq}(\phi^+_0) 
  \nonumber \\
           &=& \frac{1}{(a^+)^2} \left\{ T \left[ \phi^+_0\ln\phi^+_0 
               + (1-\phi^+_0)\ln(1-\phi^+_0) - \phi^+_0\ln\phi^+_b - 
         \frac{2}{\lambda}(\sqrt{(\lambda\phi^+_0)^2+1} - 1) \right. \right. 
               \nonumber \\ 
               & & + \left. \left. 2\phi^+_0\sinh^{-1}(\lambda\phi^+_0) \right] 
         - \alpha\phi^+_0 - \frac{\beta}{2}(\phi_0^+)^2 \right\}.
 \label{f0kinetic}
\end{eqnarray}
Expression (\ref{f0kinetic}) determines the behavior of the dynamic
surface tension as the surfactant ions adsorb at the interface and
the surface coverage increases.
Note, that it is very different from the equilibrium relation (\ref{eqstate2}),
and hence, using the equilibrium equation of state to relate $\Delta\gamma$
to $\phi^+_0$ is invalid in this case.

Assuming a strong electric field ($\lambda\phi^+_0 \gg 1$), the function
$\Delta\gamma(\phi^+_0)$ can be shown to depart from the convex shape of 
a simple well, if $\beta/T > 2(2+\sqrt{3}) \simeq 7.5$. 
For such high values of $\beta$ we should expect, therefore,
an interesting time dependence of the dynamic surface tension, as
demonstrated in Fig.~6.
It should be stressed, that the curve of Fig.~6a is not 
presumed to exactly correspond to the experimental results 
reproduced in Fig.~6b. 
Our claim is that an interfacial free energy of the form suggested by
Eq.~(\ref{f0kinetic}) may clearly account for the unusual, experimentally
observed dynamic surface tension depicted in Fig.~6b.

For values of parameters other than those chosen in 
Fig.~6, the interfacial free energy may have a 
non-monotonic, double-well shape.
In such a case, we still expect the time dependence of the surface tension
to schematically resemble the curve in Fig.~6b,
{\em i.e.}, to exhibit a period of almost constant tension as the system 
undergoes the transition from the first well to the second.
Our current, diffusive formalism, however, cannot quantitatively describe the 
kinetics of such a process \cite{Langer,unlikebarrier}.

%-----------------------------------------------------------
\vspace{0.5cm}
\subsection{The Effect of Counter-Ion Adsorption}

An issue still to be addressed is whether such high
values for the interaction constant, $\beta$, are reasonable 
for ionic surfactants.
Measurements on non-ionic surfactants yield typical values for 
$\beta$ which are smaller than $4T$ \cite{Lin2}. 
There is no obvious reason why ionic surfactants should exhibit,
apart from the Coulombic repulsion, also much stronger lateral 
attraction at the interface.
The answer lies in one of the assumptions we have used.
Throughout the analysis above, it  was assumed that
the counter-ions are surface-inactive and hence not present at all
at the interface itself. 
In fact, this assumption is inaccurate and was taken merely for
the sake of a clearer and simpler discussion.
The high surface potentials involved in salt-free surfactant 
solutions should attract a certain amount of counter-ions to 
the interface.
Our aim now is to calculate the correction introduced
into the interfacial free energy as a result of the presence of a 
small amount of counter-ions at the interface.

For simplicity, we assume the counter-ions to be in 
quasi-equilibrium throughout the system, and therefore obey the 
Boltzmann distribution (\ref{Boltzmann}),
\[
  \phi^-_0 = \phi^-_b \e^{e\psi_0/T}.
\]
Since the surface charge consists now of both surfactant ions and
counter-ions, expression (\ref{psi0}) for the surface potential
should be modified,
\begin{equation}
  \psi_0 = \frac{2T}{e} \sinh^{-1}\{\lambda[\phi^+_0-(a^+/a^-)^2\phi^-_0]\}.
 \label{A2}
\end{equation}
In the limits of a strong field ($\lambda\phi^+_0 \gg 1$) and a small 
density of adsorbed counter-ions [$\phi^-_0/(a^-)^2 \ll \phi^+_0/(a^+)^2$], 
we find from Eq.~(\ref{A2}) that the counter-ion 
surface coverage depends quadratically on the surfactant one,
\begin{equation}
  \phi^-_0 \simeq 4\lambda^2\phi^-_b (\phi^+_0)^2.
 \label{pm0}
\end{equation}

The adsorbed counter-ions introduce the following direct correction
to the interfacial free energy as was formulated in 
Sec.~\ref{ionicnosalt},
\[
  \delta(\Delta\gamma)_{counter} = \frac{1}{(a^-)^2} [ T 
         (\phi^-_0\ln\phi^-_0 - \phi^-_0) - e\phi^-_0\psi_0 -
         \mu^-_1\phi^-_0 ].
\]
In quasi-equilibrium this expression reduces to
\[
  \delta(\Delta\gamma)_{counter} = -\frac{T}{(a^-)^2} \phi^-_0,
\]
so the correction to $(a^+)^2\Delta\gamma/T$ takes the form
\begin{equation}
  \delta \left( \frac{(a^+)^2\Delta\gamma}{T} \right)_{counter} = 
         -\left( 2\lambda \frac{a^+}{a^-} \right)^2 \phi^-_b (\phi^+_0)^2.
 \label{countercorrection}
\end{equation}
In addition, the adsorbed counter-ions introduce {\em indirect} corrections 
into the free energy, through the terms $e\phi^+_0\psi_0$ and $f_{bulk}^{eq}$ 
of Eq.~(\ref{f0kinetic}).
However, these two corrections turn out to exactly cancel each other.
Note that apart from electrostatic interactions, we have neglected any
other, short-range interactions between surfactant ions and counter-ions at
the interface.

Looking at expression (\ref{countercorrection}), we find that the effect
of a small amount of adsorbed counter-ions is equivalent to an effective
increase in the lateral attraction term, $-\beta(\phi^+_0)^2/2$ 
of Eq.~(\ref{f02}).
The coefficient $(2\lambda a^+/a^-)^2\phi^-_b$ can be also written as
$2\pi l a^-/(a^+)^2$. 
Taking $a^- \leq a^+ \sim l$ leads to a correction of order 1--10, \ie,
the counter-ion adsorption may introduce, indeed, a significant
addition to the attraction term.
Note, that the increase in $\beta$ is sensitive to molecular details
(the dimensions of the two types of ions).
This fact may be related to the experimental observation that some ionic 
surfactants do exhibit the unusual time dependence depicted in 
Fig.~6b, while others do not \cite{private}.

%---------------------------------------------------------------
% Ionic Surfactants with Salt
%---------------------------------------------------------------
\section{Ionic Surfactants with Added Salt}
\label{ionicwithsalt}
\setcounter{equation}{0}

In many practical cases, the solution contains, in addition to the 
surfactant ions and their counter-ions, also a certain amount of
dissolved salt.
The effect of adding mobile ions, whose concentration usually 
exceeds that of the surfactant, is to screen the electrostatic 
interactions.
Since it was found in Sec.~\ref{ionicnosalt} that strong electrostatic
interactions drastically affect the nature of the adsorption process, 
we should expect the results in the presence of added salt to be
significantly different from the case of salt-free ionic solutions.

In principle, adding salt introduces two additional degrees of 
freedom into our formalism, namely the profiles of
the salt ions and their counter-ions.
This should significantly complicate the already elaborate
problem of ionic surfactant adsorption.
In order to avoid such complications we adopt the following 
simplifying assumptions:
\begin{itemize}
 \item The salt is surface-inactive and its concentration is 
          much larger than that of the surfactant. 
           This assumption allows us to separate the roles played by the
           two types of ions: the surfactant ions adsorb at the 
           interface and build up the surface charge, 
           whereas the salt ions form the electric double
           layer inside the solution.
 \item The salt ions are much more mobile than the surfactant ones.
           Relying on this assumption, the kinetics of the salt ions can be 
           neglected, and the double layer they form can be assumed to 
           maintain quasi-equilibrium with the changing surface charge.
 \item For simplicity, we assume that the salt ions and counter-ions are 
          monovalent as well, or more generally, have the same valency
          \cite{nosuperscript}.
\end{itemize}

Most of the formulation given in Sec.~\ref{ionicnosalt} remains valid,
in particular the equations governing the surfactant kinetics,
Eqs. (\ref{diffusion2})--(\ref{dp0dt2}).
The main difference is that the electric potential, $\psi$, is no longer 
regarded as an independent degree of freedom coupled to the surfactant 
profile, $\phi$.
Instead, it is simply given by the potential of an equilibrium double
layer, depending on the surface coverage and salt bulk concentration
alone.
The high salt concentration we have assumed, allows us to take a screened
double-layer potential in the linear, Debye-H\"{u}ckel regime 
\cite{AndelmanES},
\begin{equation}
  \psi(x,t) = \frac{4\pi e}{\epsilon\kappa a^2} \phi_0(t) \e^{-\kappa x},
\end{equation}
bearing in mind that in this section, the salt bulk concentration, $c_s$, 
replaces the surfactant one, $c_b$, in the definition of $\kappa$.
Substituting this potential in Eqs.~(\ref{diffusion2}) and (\ref{dp1dt2}),
the equations determining the kinetics inside the solution are obtained,
\begin{eqnarray}
  \frac{\pd\phi}{\pd t} &=& D \frac{\pd}{\pd x} \left(
         \frac{\pd\phi}{\pd x} - \frac{\kappa^2}{2c_s a^2}\phi_0
         \e^{-\kappa x} \phi \right),
  \label{diffusion3} \\
  \frac{\pd\phi_1}{\pd t} &=& \frac{D}{a} \left( \left. \frac{\pd\phi}{\pd x} 
         \right|_{x=a} - \frac{\kappa^2}{2c_s a^2}\phi_0\phi_1 \right) -
         \frac{\pd\phi_0}{\pd t}.
  \label{dp1dt3}
\end{eqnarray}
In the last equation we have also assumed that the Debye-H\"{u}ckel 
screening length is much larger than the surfactant molecular dimension, 
$\kappa a \ll 1$.
The equation describing the adsorption kinetics at the interface, 
Eq.~(\ref{dp0dt2}), remains valid as it is.

As in the previous two sections, we are interested in the distinction 
between the two limits --- diffusion-limited vs. kinetically
limited adsorption.
In order to find the time scale of diffusion, we treat  the electric field as a 
small perturbation and seek a solution to Eqs. (\ref{diffusion3}) and 
(\ref{dp1dt3}) which is close to the one in the non-ionic case.
A procedure similar to the one given in Sec.~\ref{non-ionic} is now
employed, yielding the following asymptotic expression,
\begin{equation}
  \phi_b - \phi_1 \simeq \frac{\kappa\phi_b\phi_{0,eq}}{2c_s a^2}
       + \frac{a\phi_{0,eq}}{\sqrt{\pi D t}} \left[ 1 - \frac{c_b}{2c_s} -
       \frac{\kappa\phi_{0,eq}}{2c_s a^2} \left( 1 - \frac{3c_b}{2c_s}
       \right) \right].
 \label{asymp3}
\end{equation}
When this result is compared to its non-ionic parallel, Eq.~(\ref{asymp}),
two observations are to be made.
The first is that Eq.~(\ref{asymp3}) contains an additional constant term,
since the equilibrium sub-surface concentration differs from
the bulk one in the presence of an electric field.
The second is that the diffusion time scale is slightly corrected,
as expected, by the weak electrostatic interactions,
\begin{equation}
  \tau_d = \tau_d^{(0)} \left[ 1 - \frac{c_b}{2c_s} - \frac{\kappa\phi_{0,eq}}
          {2c_s a^2} \left( 1 - \frac{3c_b}{2c_s} \right) \right]^2,
\end{equation}
where $\tau_d^{(0)}$ denotes the diffusion time scale found in the non-ionic case
[Eq.~(\ref{td})].
The correction vanishes when we take very high salt concentrations,
leading to a complete screening of electrostatic interactions.
            
Turning to the case of KLA, we find that the
expression for the time scale of kinetics at the interface derived for
salt-free solutions, Eq.~(\ref{tk2}), remains valid also in the
presence of added salt.
However, unlike the case of Sec.~\ref{ionicnosalt}, the electric field 
in this case is weak, and the resulting slowing factor is small, so that
$\tau_k \geq \tau_k^{(0)}$.
Since the time scales of both the diffusion inside the solution and 
the kinetics at the interface have been shown, in the presence of salt,
to differ only slightly from those of the non-ionic case, we conclude that
{\em ionic surfactants with added salt, like non-ionic surfactants,
 should exhibit diffusion-limited adsorption}.
As in the case of non-ionic surfactants, this conclusion is well
supported by experiments, as illustrated in Fig.~7.
Indeed, measurements on ionic surfactant solutions with salt can be 
well fitted by theoretical curves,  using the same schemes used
for non-ionic surfactants (see Sec.~\ref{non-ionic}) \cite{Langevin}.
         
%---------------------------------------------------------------
% Concluding Remarks
%---------------------------------------------------------------
\section{Concluding Remarks}
\label{conclusion}
\setcounter{equation}{0}

In this work we have presented an alternative approach to the 
problem of the kinetics of surfactant adsorption.
One of the advantages of this approach is that the diffusion inside 
the aqueous solution and the kinetics of adsorption at the interface 
are not introduced as two separate, independent processes, but both 
arise from the same model.
This makes the model more complete than previous ones,
 and allows us to point at the process limiting the kinetics of 
the entire system in various cases.
We find the adsorption to be limited by the bulk diffusion
in the cases of non-ionic surfactants and ionic surfactants with added
salt, and by the kinetics at the interface in the case of salt-free
ionic surfactant solutions.
Such conclusions lead to a significant mathematical simplification of the 
statement of the problem. 
They are also in agreement with experimental findings.

Another advantage is that the formulation can be readily extended
to more complicated systems.
We have used this to account for the kinetic adsorption of ionic 
surfactants with and without added salt.
In particular, we have been able to explain the recently reported unusual 
time dependence of the surface tension in salt-free ionic surfactant 
solutions.
In addition, our free-energy approach provides a general, straightforward
method for calculating dynamic surface tensions from surface coverages, 
which does not rely on equilibrium relations.
This feature turns out to be essential in the case of salt-free ionic 
surfactant solutions.

The adsorption of ionic surfactants behaves very differently in the 
presence or absence of salt: it is diffusion-limited in the former
case, and kinetically limited in the latter case.
This has been shown both experimentally and by our theory.
In order to reach better understanding of the kinetics of surfactant
adsorption, additional experiments are required, particularly on
ionic surfactants.
All experiments until now have involved aqueous solutions which are 
either free of salt or containing high concentrations of it.
It should be interesting to examine aqueous solutions with low salt  
concentrations and observe the crossover from one limiting
behavior to another.
Moreover, since the adsorption in the case of salt-free ionic surfactant
solutions has been found to be kinetically limited, dynamic surface
tension measurements may be used to ``probe'' the actual dependence of
the interfacial free energy, $f_0(\phi_0)$, on the surface coverage,
$\phi_0$. 
This, in turn, may help explain the equilibrium phase behavior of 
ionic surfactant monolayers under various conditions 
(\eg, under compression).

The theory presented in this work is incomplete in two main
aspects.
The first is that our diffusive formalism,  as was mentioned in 
Sec.~\ref{ionicnosalt}, cannot fully describe the kinetics in cases where an 
energy barrier must be overcome before the interfacial free energy reaches its 
minimum.
According to the analysis of Sec.~\ref{ionicnosalt}, actual dynamic surface tension
measurements imply that the free energy in the case of certain ionic 
surfactant solutions probably does exhibit such a barrier.
Quantitative treatment of the evolution of such systems should require, therefore,
a more accurate (perhaps ``Kramers-like'' \cite{Kramers}) theory.
The second aspect is the lateral diffusion at the interface, whose time scale
has been completely neglected by this work.
As was mentioned in Sec.~\ref{non-ionic}, cases where lateral diffusion is
significant are encountered in practice, and a future, more complete theory
cannot ignore its effect.

%------------------------------------------------------------
% Acknowledgments
%------------------------------------------------------------
\vspace{3cm}
\newlength{\tmp}
\setlength{\tmp}{\parindent}
\setlength{\parindent}{0pt}
{\em Acknowledgments}
\setlength{\parindent}{\tmp}

We are indebted to D.~Langevin and A.~Bonfillon-Colin for introducing
us to the problem of dynamic surface tension, sharing with us their 
unpublished experimental data and for illuminating discussions.
We also benefited from discussions with N.~Agmon.
Support from the German-Israeli Foundation (G.I.F.)
under grant No.~I-0197 and the US-Israel Binational Foundation (B.S.F.)
under grant No.~94-00291 is gratefully acknowledged.

%------------------------------------------------------------
% References
%------------------------------------------------------------

%------------------------------------------------------------
% Figure Captions
%------------------------------------------------------------
\pagebreak
\section*{Figure Captions}

\begin{itemize}

 \item[{\bf Fig.~1}] 
  Schematic view of the system. A sharp, flat interface separates
  a dilute aqueous solution of non-ionic surfactants from an air or
  oil phase.

 \item[{\bf Fig.~2}]
  The two limiting cases for the time dependence of the adsorption:
  (a) Diffusion-limited adsorption (DLA) --- the surface coverage, $\phi_0$,
     is determined by the minimum of the surface free energy, $f_0$,
     yet the shape of $f_0$ changes with time. The curve is shown
     at three different times, corresponding to increasing values
     of the sub-surface volume fraction, $\phi_1$: (i) at $t_1$ close 
     to the beginning of the process, when $\phi_1=5\times 10^{-7}$,
     (ii) at a later time $t_2$, when $\phi_1=1.5\times 10^{-6}$ and 
     (iii) at equilibrium, when $\phi_1=\phi_b=5\times 10^{-6}$.
     The energy constants are set to the (realistic) values
     $\alpha=12T$ and $\beta=3T$;
  (b) Kinetically limited adsorption (KLA) --- the shape of $f_0$ is fixed and
     $\phi_0$ increases with time until reaching equilibrium at the 
     minimum of $f_0$.

 \item[{\bf Fig.~3}]
   A variety of non-ionic surfactants were experimentally 
   found to exhibit DLA. 
   Four examples of dynamic surface tension measurements are given:
   $9.49\times 10^{-5}$M of decyl alcohol (open circles),
   as adapted from Ref.~\cite{AddHutch};
   $2.32\times 10^{-5}$M of Triton X-100 (squares),
   adapted from Ref.~\cite{Lin1};
   $6\times 10^{-5}$M of C$_{12}$EO$_8$ (triangles)
   and $4.35\times 10^{-4}$M of C$_{10}$PY (solid circles),
   both adapted from Ref.~\cite{Hua2}.
   Note the asymptotic $t^{-1/2}$ behavior, characteristic
   of DLA and shown by the solid fitting lines.

 \item[{\bf Fig.~4}]
  (a) The dependence between the surface tension and surface coverage in
     a DLA case (the parameters are the same as in Fig.~2);
  (b) The schematic time dependence of the surface tension expected
      from a dependence $\Delta\gamma(\phi_0)$ such as in (a):
      an initial slow change, then a rapid drop and finally a relaxation 
      towards equilibrium.

 \item[{\bf Fig.~5}]
  Schematic view of the ionic salt-free system. 
  A dilute solution containing surfactant ions and their counter-ions 
  has a sharp, flat interface with an air or oil phase.

 \item[{\bf Fig.~6}]
  (a) The dependence between surface tension and surface 
     coverage in the KLA case, according to Eq.~(\ref{f0kinetic}), 
     where the values taken for the parameters are: 
     $a^+=12$ \AA, $\phi^+_b=3.6\times 10^{-4}$, $\alpha=11.5T$ and 
     $\beta=9T$;
  (b) The experimentally observed dynamic surface tension of a 
      salt-free $3.5\times 10^{-4}$M SDS solution against dodecane, 
      adapted from Ref.~\cite{Langevin}. 
      The time dependence is schematically consistent with a non-convex
      dependence $\Delta\gamma(\phi^+_0)$ such as the one in (a).

 \item[{\bf Fig.~7}]
   Ionic surfactants in the presence of added salt
   were experimentally found to exhibit DLA. 
   Three examples are given:
   Dynamic surface tension of $4.86\times 10^{-5}$M SDS with 0.1M 
   NaCl against dodecane (open circles and left ordinate),
   adapted from Ref.~\cite{Langevin};
   Dynamic surface tension of $2.0\times 10^{-4}$M SDS with 0.5M
   NaCl against air (squares and left ordinate),
   adapted from Ref.~\cite{Fainerman};
   Surface coverage deduced from Second Harmonic Generation 
   measurements on a saturated solution of SDNS with 2\% NaCl against
   air (solid circles and right ordinate), adapted from Ref.~\cite{SHG}.
   Note the asymptotic $t^{-1/2}$ behavior, characteristic
   of DLA and shown by the solid fitting lines.

\end{itemize}

%------------------------------------------------------------
% Figures
%------------------------------------------------------------

\pagebreak

\begin{figure}[p] 
\epsfysize=11\baselineskip
\centerline{\hbox{\epsffile{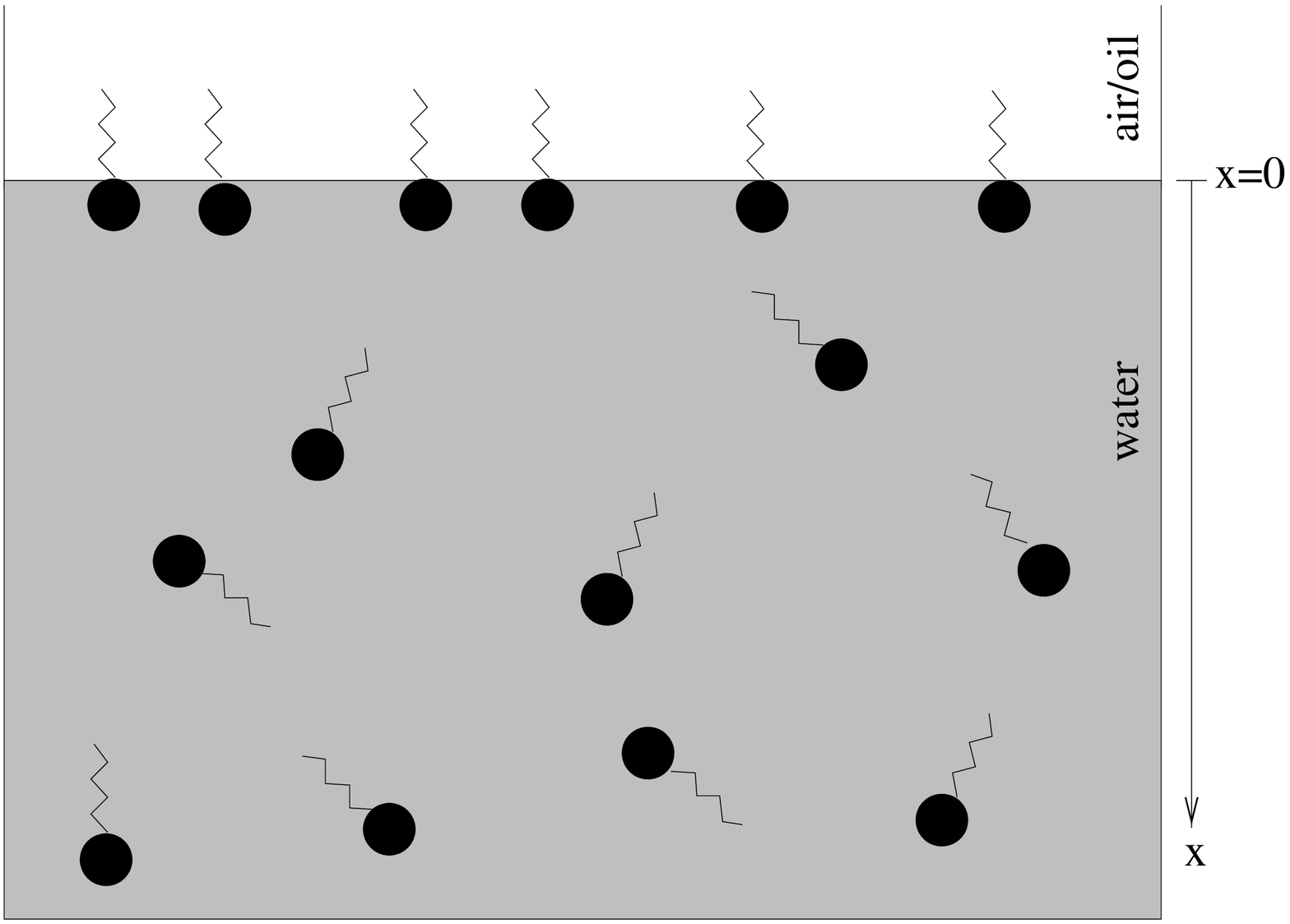}}} 
\caption[]{}
\label{sysscheme1}
\end{figure}

\begin{figure}[p] 
\epsfysize=16\baselineskip
\centerline{\hbox{\epsffile{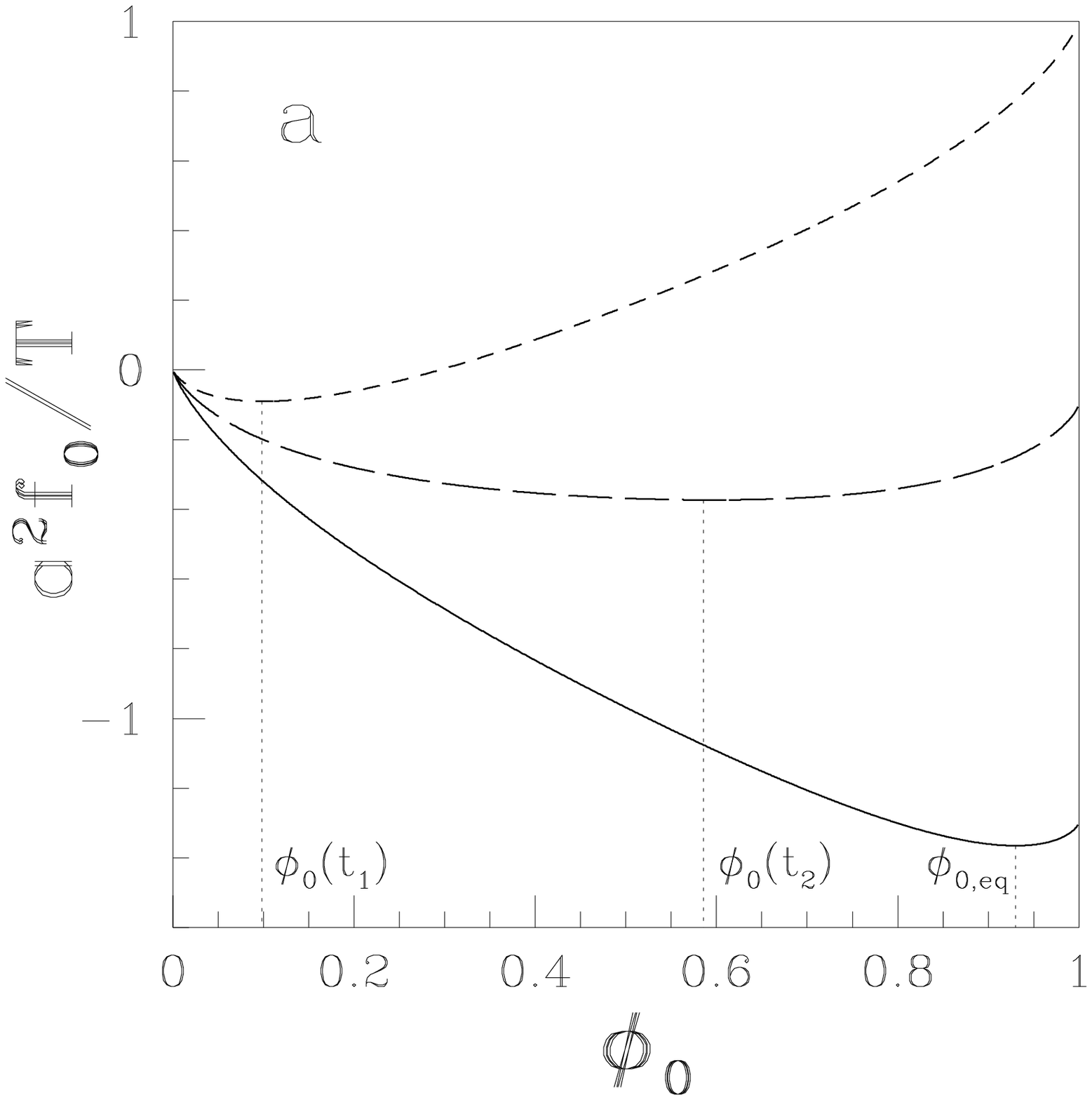}} 
\epsfysize=16\baselineskip \hbox{\epsffile{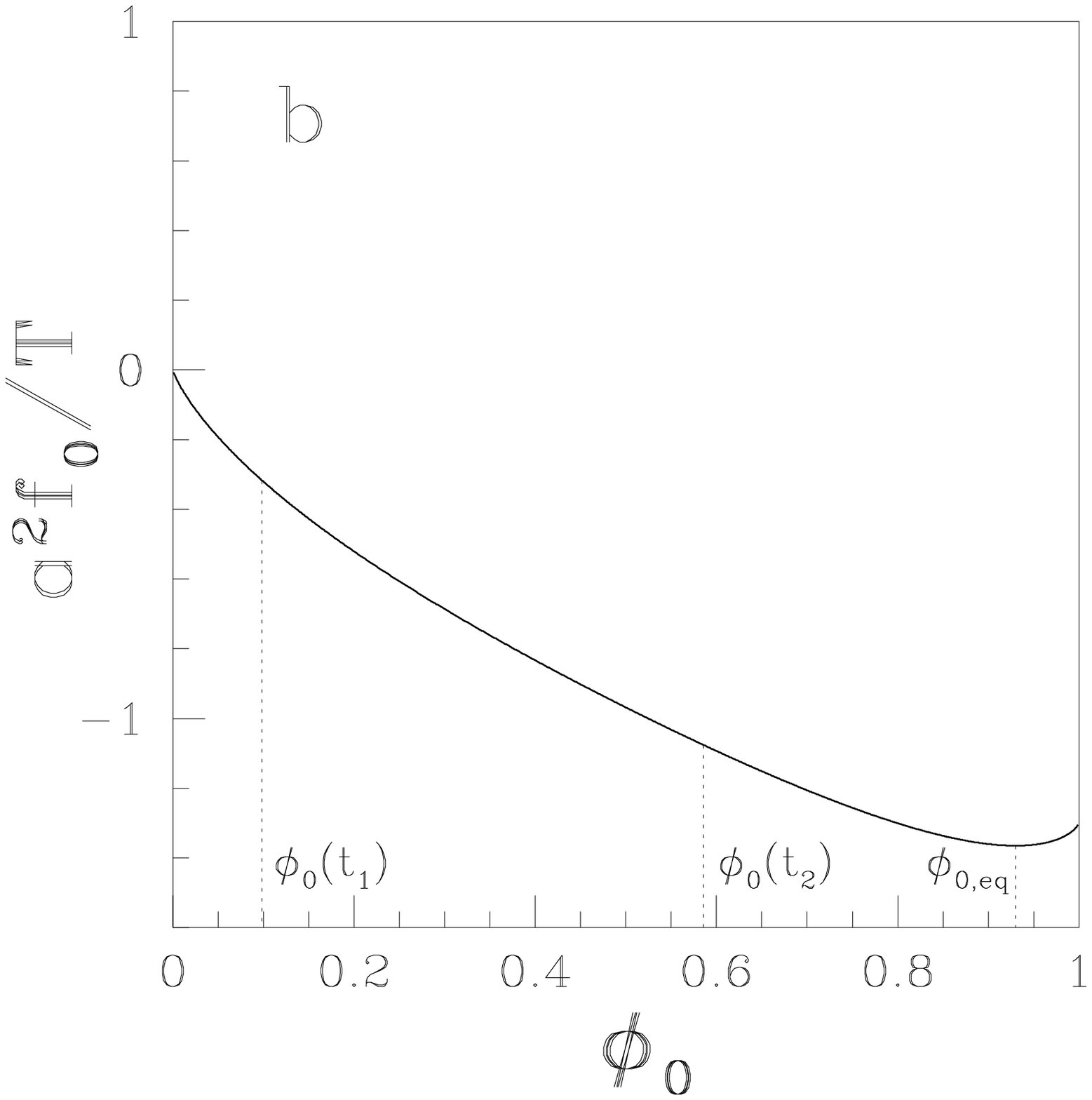}}}
\caption[]{}
\label{adstype}
\end{figure}

\begin{figure}[p] 
\epsfysize=16\baselineskip
\centerline{\hbox{ \epsffile{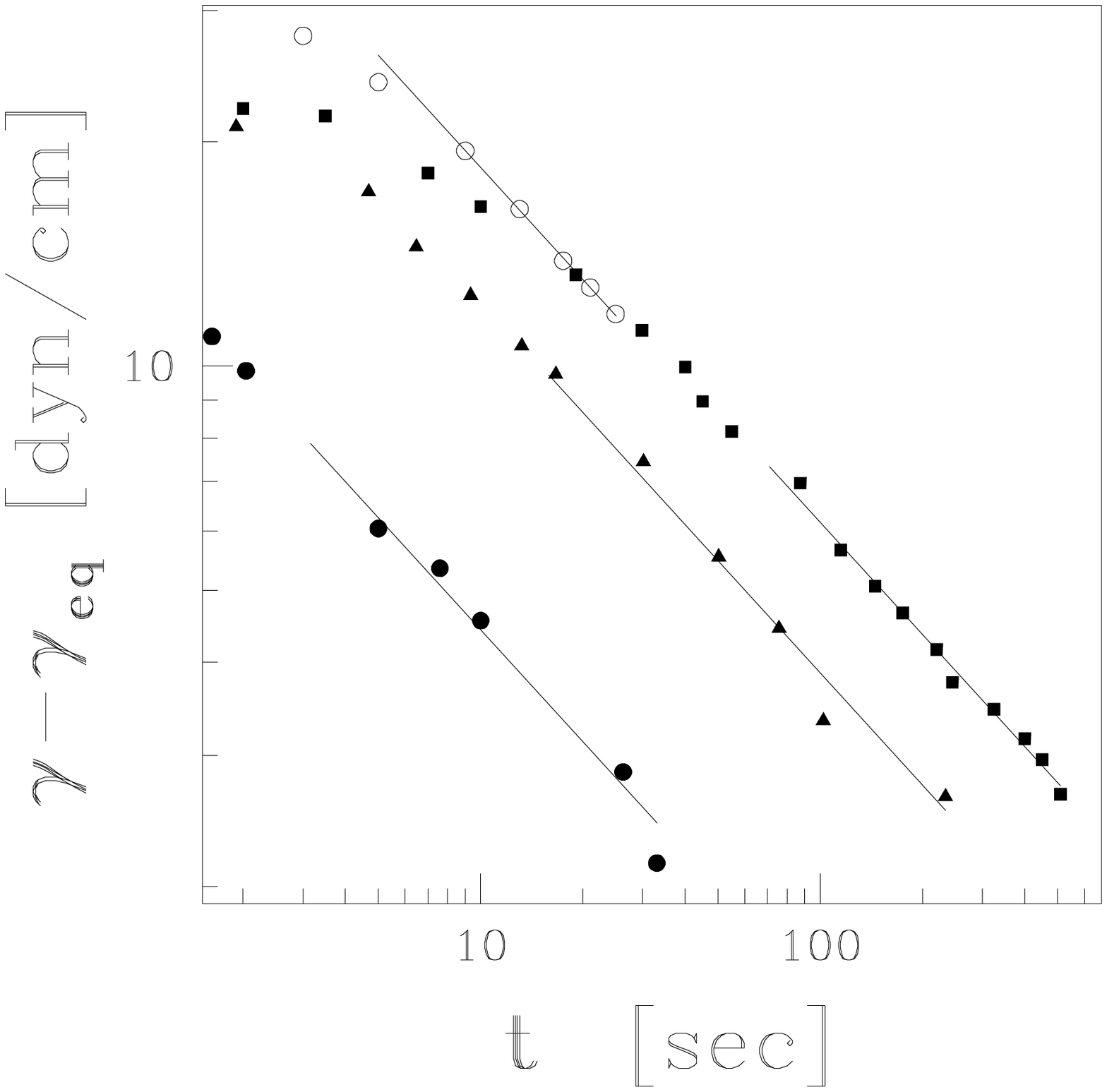} }} 
\caption[]{}
\label{diffcntl}
\end{figure}

\begin{figure}[p] 
\epsfysize=16\baselineskip
\centerline{\hbox{\epsffile{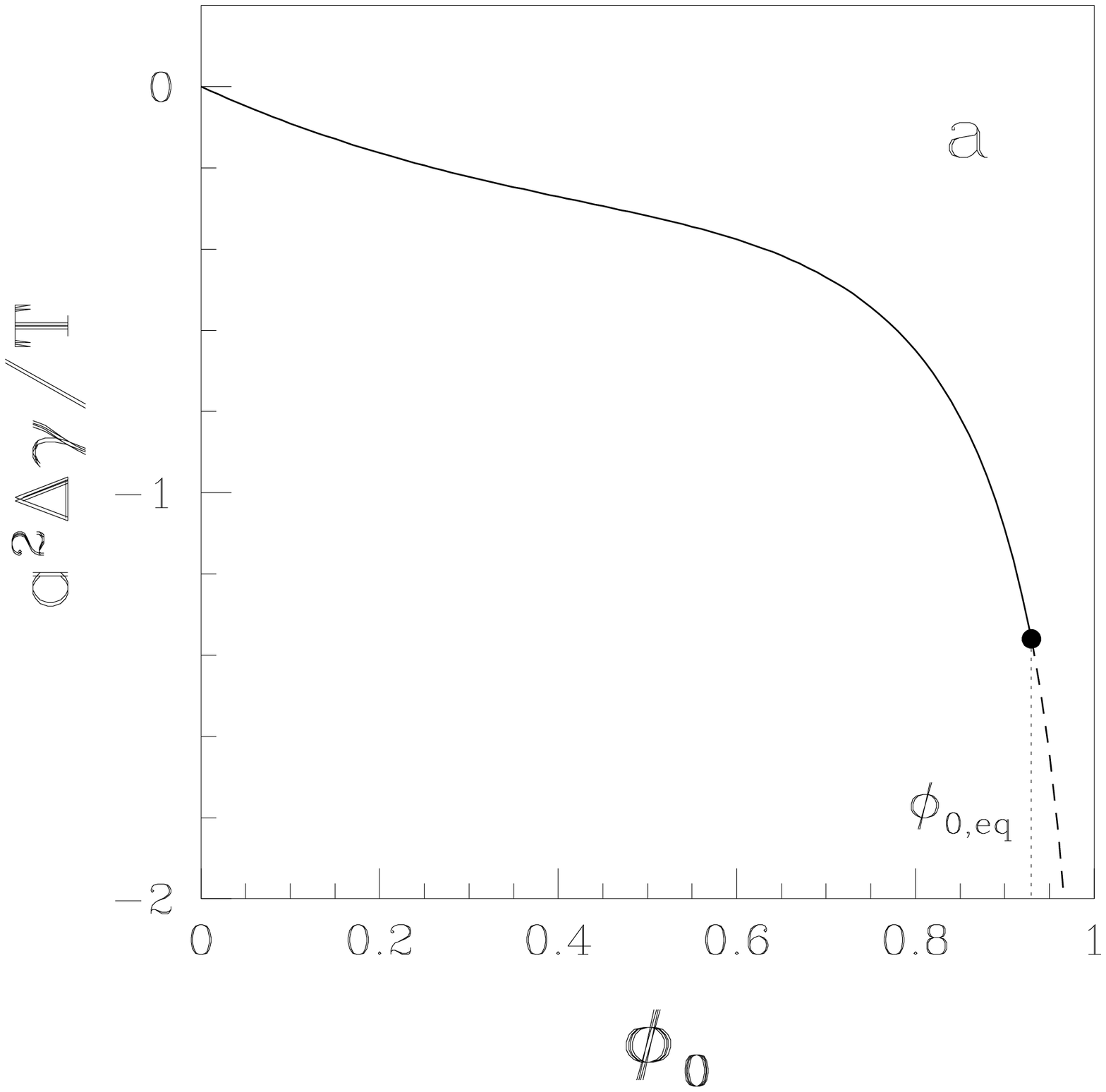}} 
\epsfysize=16\baselineskip \hbox{\epsffile{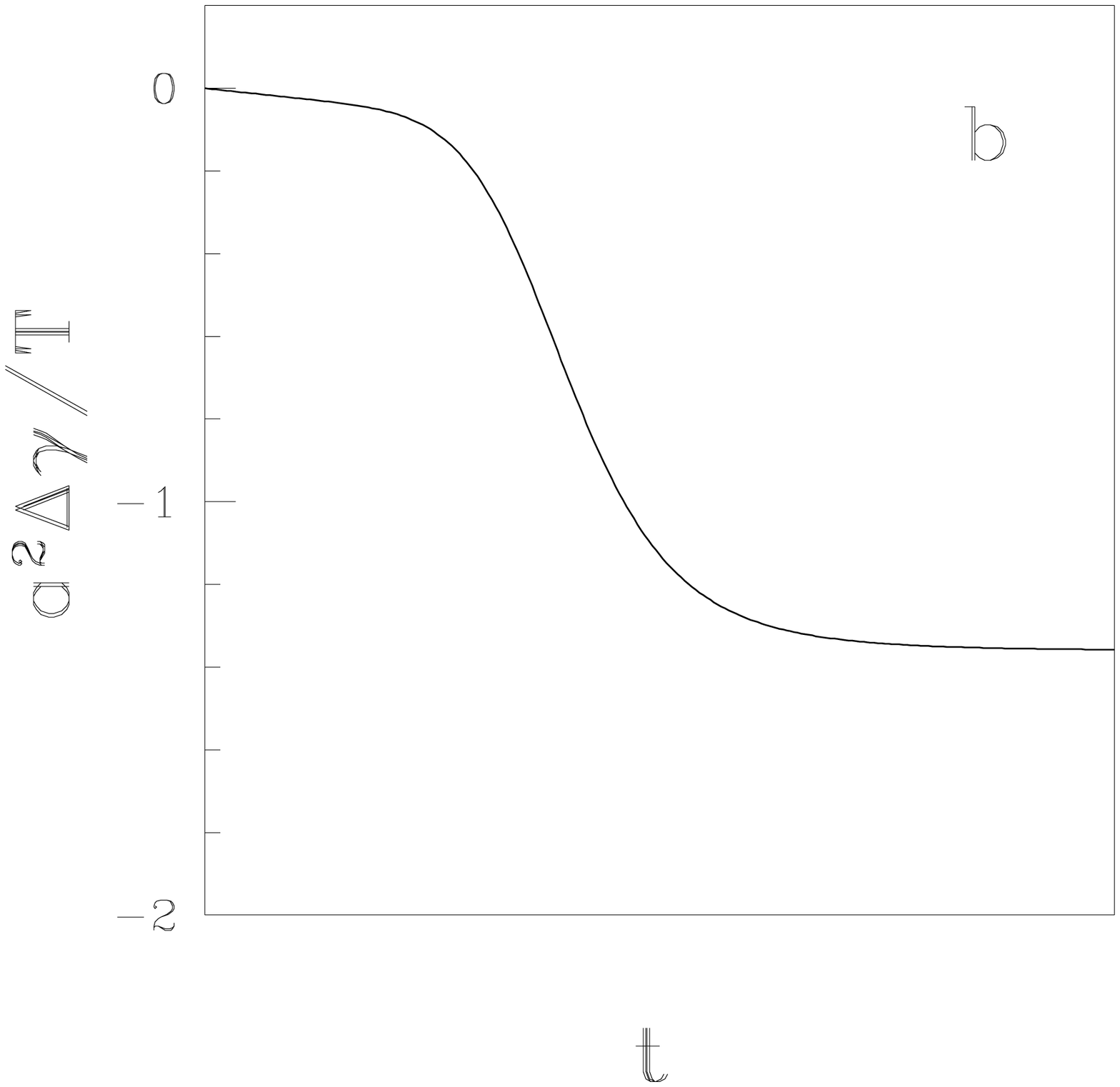}}}
\caption[]{}
\label{qualitative}
\end{figure}  

\begin{figure}[p] 
\epsfysize=11\baselineskip
\centerline{\hbox{\epsffile{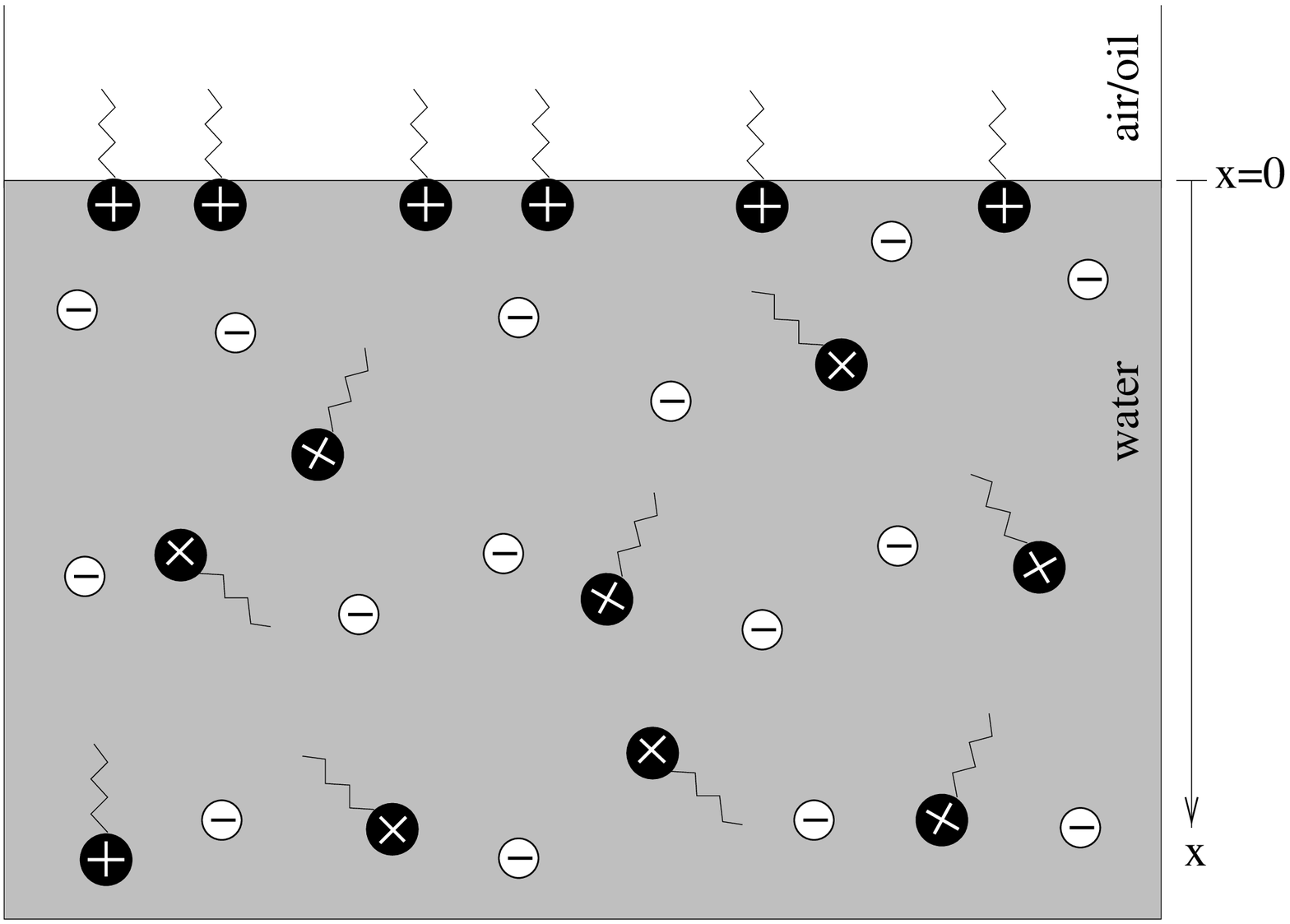}}} 
\caption[]{}
\label{sysscheme2}
\end{figure}

\begin{figure}[p] 
\epsfysize=16\baselineskip
\centerline{\hbox{\epsffile{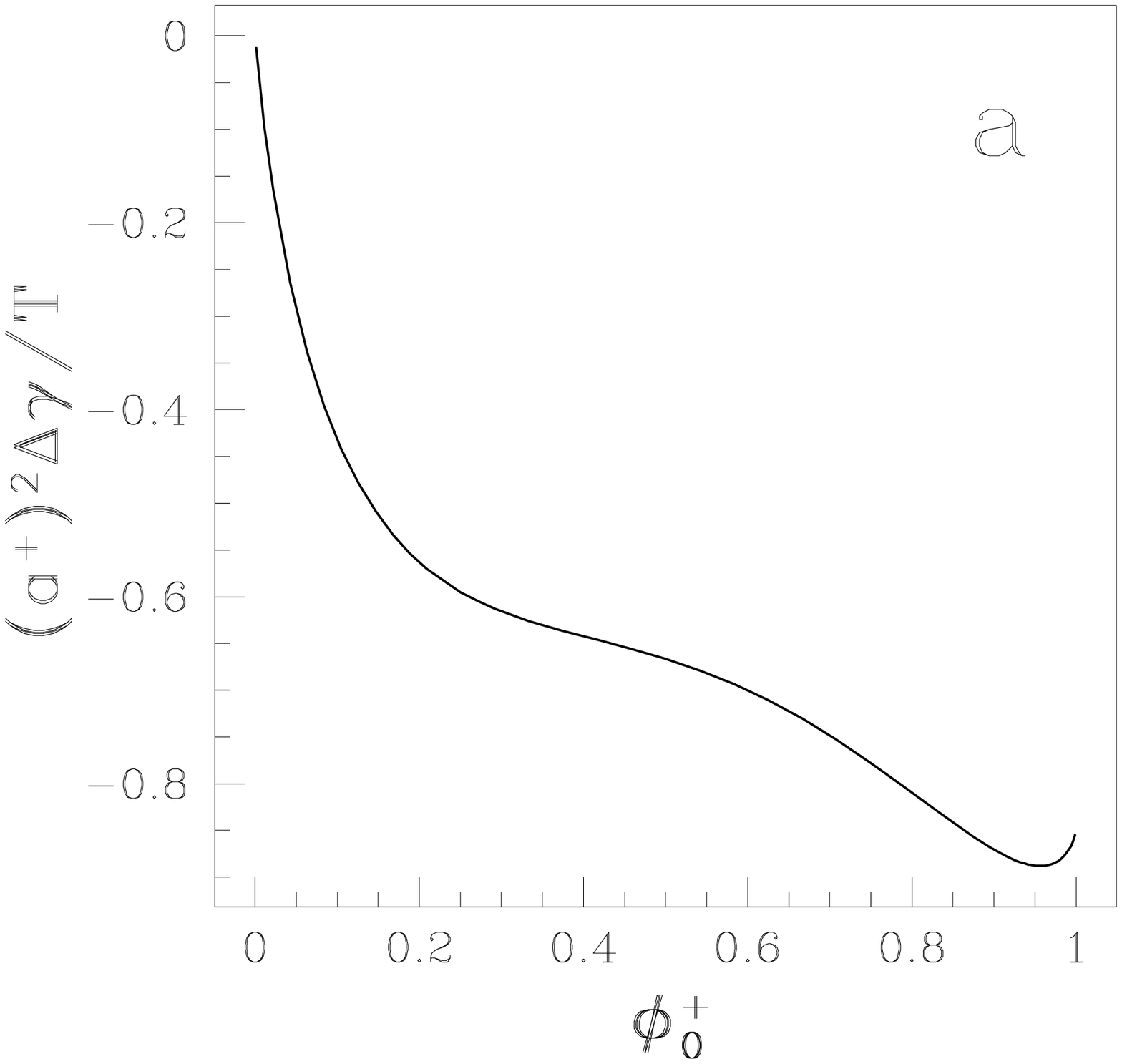}} 
\epsfysize=16\baselineskip \hbox{\epsffile{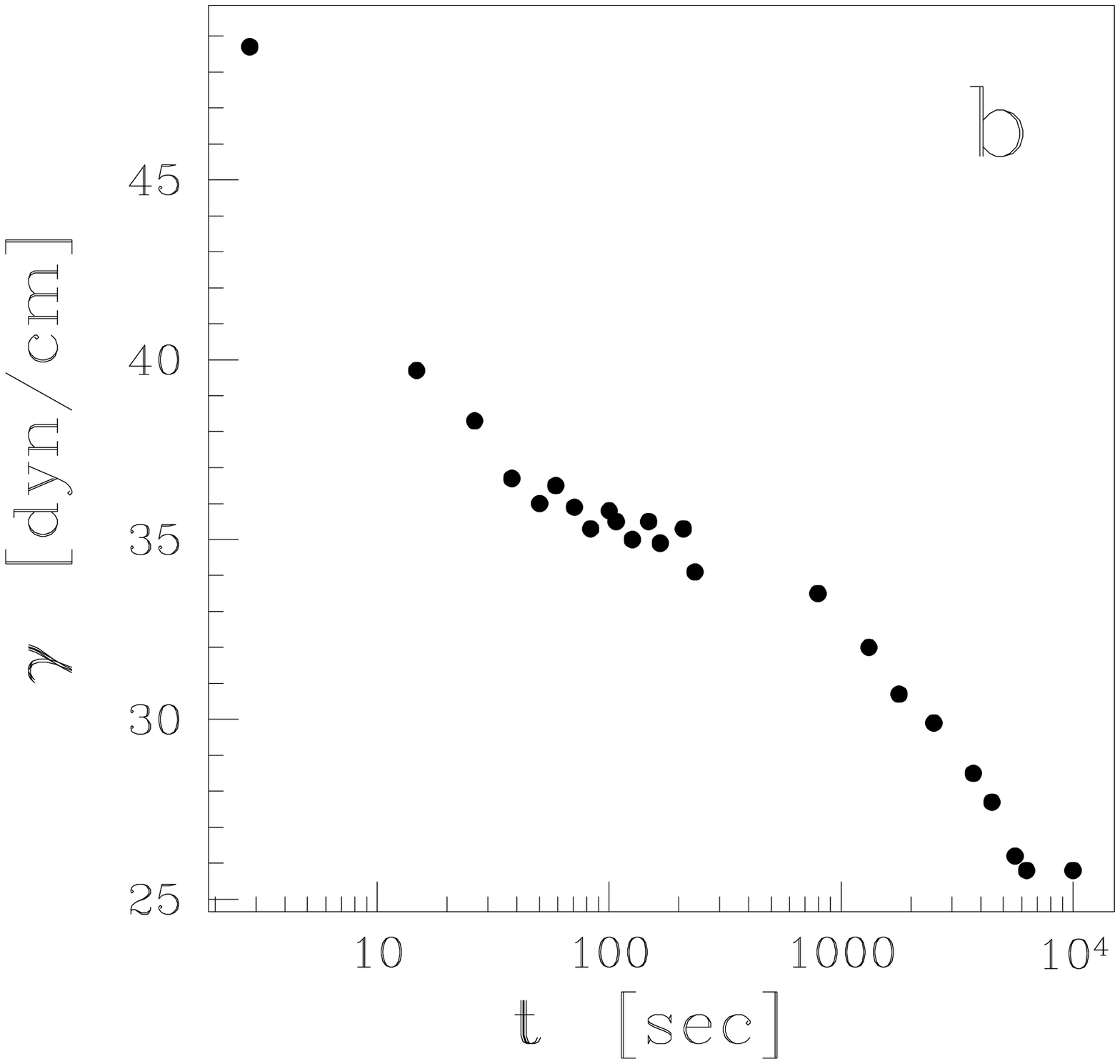}}}
\caption[]{}
\label{plateau}
\end{figure}

\begin{figure}[p] 
\epsfysize=16\baselineskip
\centerline{\hbox{ \epsffile{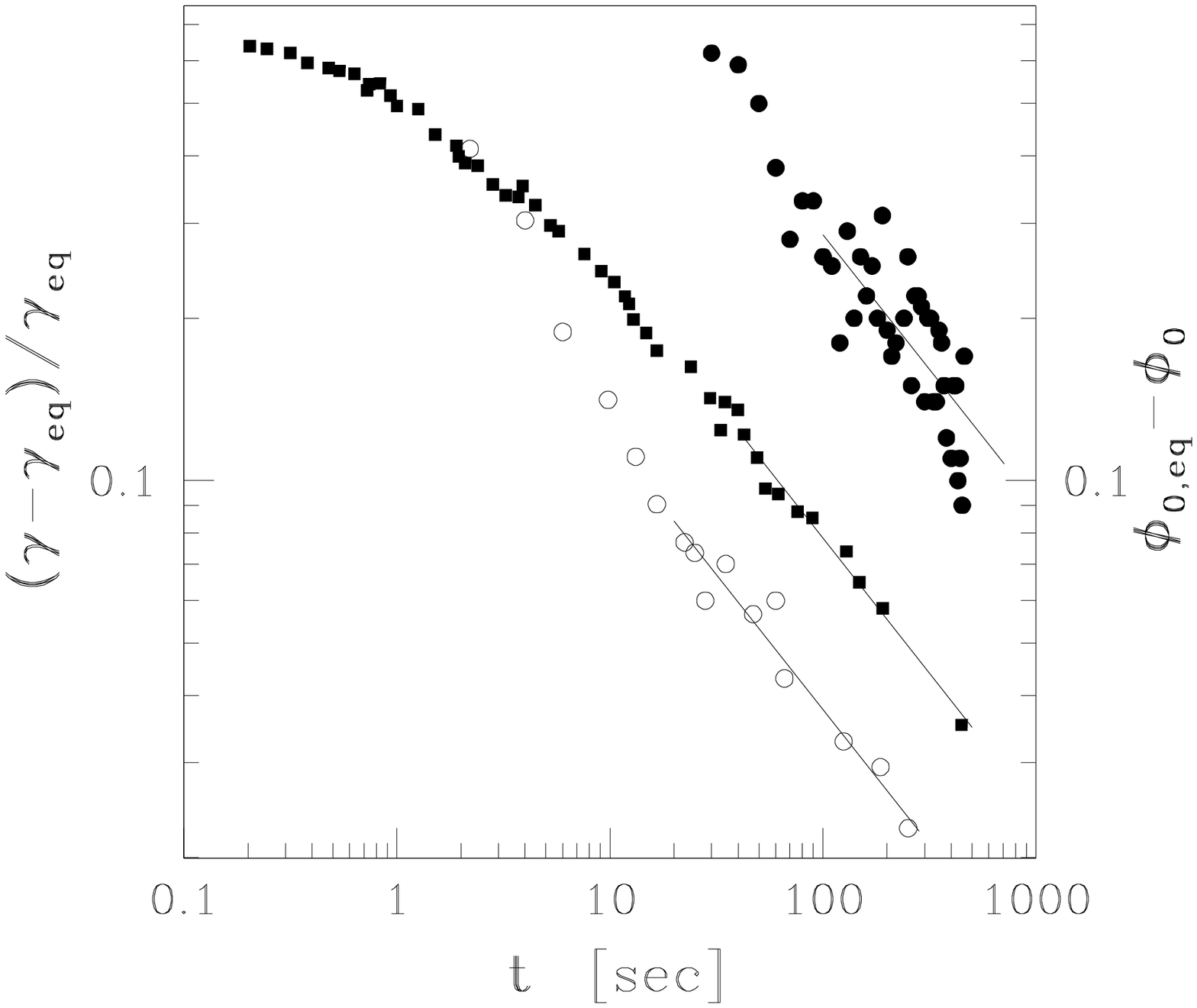} }} 
\caption[]{}
\label{diffcntlsalt}
\end{figure}

%------------------------------------------------------------


\begin{thebibliography}{99}
\bibitem{barrier} Note, that we did not include in expression (\ref{f0}) a 
potential barrier hindering the surfactant from adsorbing at the interface. 
The introduction of such a barrier amounts to adding to (\ref{f0}) a 
positive term linear in $\phi_0$, as we shall do, indeed, when 
considering the electrostatic repulsion in Sec.~\ref{ionicnosalt}.
\bibitem{differD} Note, that the conditions at the interface are 
quite different from those inside the aqueous solution, and the
diffusivities, $D$, appearing in Eqs.~(\ref{jx})--(\ref{dp0dt}), 
cannot be expected to have strictly the same value, as we have assumed
here for simplicity.
\bibitem{prevkinetic} Close to equilibrium we can also write
  \[ 
  \pd\phi_0/\pd t \simeq (D\phi_b/a^2\phi_{0,eq}) [
      \e^{(\alpha+\beta\phi_0)/T}\phi_1(1-\phi_0) - \phi_0 ],
  \]
  which coincides with the adsorption-desorption form of the Frumkin (or 
  Langmuir, when $\beta=0$) kinetic equation used by previous works
  \cite{Miller}--\cite{Franses}.
\bibitem{latdiff} Note, that in the discussion above we have
completely neglected a third time scale --- the time needed 
for lateral diffusion and molecular re-orientation at the interface. 
If, however, due to certain molecular constraints, this time scale
is no longer negligible, exceptions to the above conclusions are to 
be expected (see Refs.~\cite{Joos2,Menger}).
\bibitem{counterads} As we shall see, this assumption is not 
strictly valid, \ie, the adsorption of counter-ions at the interface 
should not be ignored.
We use it here for the sake of clarity, and shall return to this point 
later.
\bibitem{unlikebarrier} The limitation which the adsorption faces in
such a case originates from the non-convexity of the interfacial free
energy, \ie, it is a collective, {\em macroscopic} effect. 
It should not be confused with a possible potential barrier,
of {\em microscopic} origin, that the surfactant may face when trying 
to move from the bulk to the interface. 
The latter {\em can} be accounted for by our model,
as explained in Sec.~\ref{non-ionic} and demonstrated
in the current section for the case of an electrostatic barrier.
\bibitem{nosuperscript} Since these assumptions render the profiles of
the salt ions and the surfactant counter-ions irrelevant to the
analysis  of the adsorption, we omit in this section 
the superscript `+' from the variables and constants associated with the 
surfactant.
\end{thebibliography}
\end{document}